\documentclass[lettersize,journal]{IEEEtran}
\PassOptionsToPackage{hyphens}{url}

\usepackage{amsmath,amsfonts}
\usepackage{array}
\usepackage[caption=false,font=normalsize,labelfont=sf,textfont=sf]{subfig}
\usepackage{textcomp}
\usepackage{stfloats}
\usepackage{url}
\usepackage{verbatim}
\usepackage{graphicx}
\usepackage{cite}
\usepackage{hyperref}
\usepackage{enumitem}
\usepackage{tikz}
\usepackage{color}
\newcommand{\red}[1]{\textcolor{red}{#1}}

\definecolor{brown}{RGB}{139,64,0}

\newcommand{\wq}[1]{{\color{red}#1}} % comments
\usepackage{caption}
\usepackage{algorithm}
\usepackage{algorithmicx}
\usepackage{algpseudocode}
\usepackage{booktabs}
\floatname{algorithm}{Algorithm} %算法
\usepackage{multirow}

\makeatletter
\renewcommand{\maketag@@@}[1]{\hbox{\m@th\normalsize\normalfont#1}}%
\makeatother

\begin{document}

\title{Score-based Generative Diffusion Models\\ for Social Recommendations}

\author{Chengyi Liu\dag, Jiahao Zhang\dag, Shijie Wang, Wenqi Fan*, Qing Li*, ~\IEEEmembership{Fellow,~IEEE}
        % <-this % stops a space

\IEEEcompsocitemizethanks{

\IEEEcompsocthanksitem C. Liu, J. Zhang, S. Wang and Q. Li are with the Department of Computing, The Hong
Kong Polytechnic University. E-mail: chengyi.liu@connect.polyu.hk; tony-jiahao.zhang@connect.polyu.hk; shijie.wang@connect.polyu.hk; csqli@comp.polyu.edu.hk.
\IEEEcompsocthanksitem W. Fan is with the Department of Computing (COMP),  and Department of Management and Marketing (MM), The Hong Kong Polytechnic University. E-mail: wenqifan03@gmail.com.
}
\thanks{(\dag\ Both authors contributed equally to this research.)}
\thanks{(* Corresponding author: Wenqi Fan.)}
}

\markboth{Journal of \LaTeX\ Class Files,~Vol.~14, No.~8, August~2021}%
{Shell \MakeLowercase{\textit{et al.}}: A Sample Article Using IEEEtran.cls for IEEE Journals}

\IEEEpubid{0000--0000/00\$00.00~\copyright~2021 IEEE}

\maketitle

\begin{abstract}

With the prevalence of social networks on online platforms, social recommendation has become a vital technique for enhancing personalized recommendations. The effectiveness of social recommendations largely relies on the social homophily assumption, which presumes that individuals with social connections often share similar preferences. However, this foundational premise has been recently challenged due to the inherent complexity and noise present in real-world social networks.
In this paper, we tackle the low social homophily challenge from an innovative generative perspective, directly generating optimal user social representations that maximize consistency with collaborative signals. 
Specifically, we propose the Score-based Generative Model for Social Recommendation (SGSR), which effectively adapts the Stochastic Differential Equation (SDE)-based diffusion models for social recommendations. 
To better fit the recommendation context, SGSR employs a joint curriculum training strategy to mitigate challenges related to missing supervision signals and leverages self-supervised learning techniques to align knowledge across social and collaborative domains.
Extensive experiments on real-world datasets demonstrate the effectiveness of our approach in filtering redundant social information and improving recommendation performance. 
% Our codes are available at ~\url{https://github.com/Anonymous-CodeRepository/Score-based-Generative-Diffusion-Models-for-Social-Recommendations-SGSR}

\end{abstract}

\begin{IEEEkeywords} 
Recommender systems, score-based generative diffusion models, social graph denoising.
\end{IEEEkeywords}

\section{Introduction}
\label{Introduction}

% 1: Recommendation -> Sparsity of U-I graph -> social recommendation (or any story from prestigious groups)
In the era of information explosion, Recommender Systems (RS) have emerged as indispensable tools for mitigating information overload and personalizing users' content stream across e-commerce and social media platforms. Collaborative Filtering (CF), which aims to learn user and item representations from historical interactions, forms the foundation of modern recommender systems ~\cite{wang2019neural, he2020lightgcn, wang2025knowledge, wang2024graph}. 
However, CF-based methods face a persistent challenge due to their reliance on the quality of user-item interaction data: real-world interaction data used for training is often sparse, which can adversely impact the performance of these models ~\cite{idrissi2020systematic, zhao2025investigating, ning2025retrieval}.
To address the data sparseness issue, social recommendation methods incorporate social context to enhance the modelling of user intent within user representations. This practice aligns with the idea that individuals connected through social ties often exhibit similar preferences—a phenomenon supported by social influence theory~\cite{cialdini2004social} and also recognized as \emph{social homophily} ~\cite{mcpherson2001birds, jiang2024challenging}. %\cite{li2022disentangled}.
In recent years, a wide range of research has focused on enhancing the performance of social recommendations by leveraging the social homophily condition, encompassing Graph Neural Networks (GNNs) methods~\cite{fan2019graph, zhang2024graph, wang2023multi, he2025graph} and Self-Supervised Learning (SSL) methods~\cite{wu2022disentangled, xia2023disentangled}. 

\begin{figure}[t!]
\centering
\includegraphics[width=0.45\textwidth]{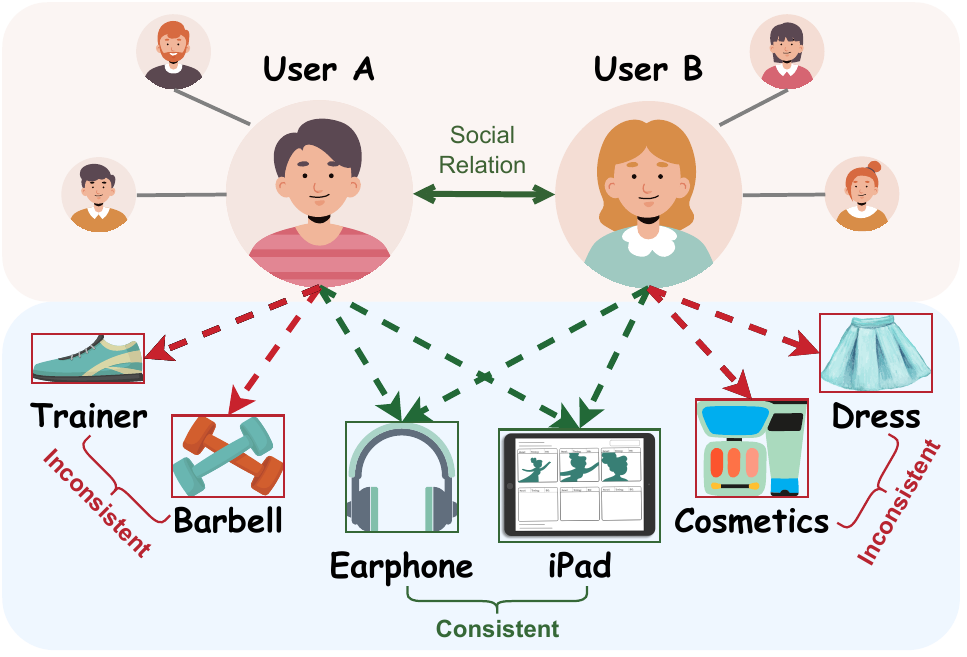}%
% \vskip -0.1in
\caption{An illustration of social connections inconsistent with user-item interactions. 
User A and User B share a social connection as classmates. 
While both exhibit interest in electronic products (\emph{social homophily}), User B's interactions with dresses and cosmetics are inconsistent with User A's preferences, introducing redundant social noise to User A's recommendation.} % This scenario illustrates how raw social graph data may potentially compromise recommendation system accuracy. }
\label{fig:so_homo} 
\vskip -0.13in
\end{figure}

\IEEEpubidadjcol

% 2: Limitation of social recommendation -- low social homophily; irrelevant information in social graph degrading recommendation performance, which necessitates social denoising for social recommendation.
Nevertheless, the underlying assumption of social homophily has recently been questioned. Extensive statistical and empirical analyses ~\cite{quan2023robust, tao2022revisiting, jiang2024challenging, yang2024graph} have suggested that leveraging the raw social graph can be unreliable and may further degrade the recommendation performance. 
This counter-intuition arises because of the inherent complexity and noise present in real-world social networks ~\cite{shu2017user}, introducing noisy information into learned user representations. Considering the example in Fig.~\ref{fig:so_homo}, a male student (i.e., User A) might establish social relationships with a female classmate (i.e., User B) who has recently acquired cosmetics or feminine apparel, yet such relationships likely contribute to unreliable recommendations for the male.
Consequently, these noisy social signals can propagate through social encoder layers, accumulating bias and ultimately hindering the recommender system's effectiveness ~\cite{tao2022revisiting}. 
Thus, it is necessary to develop social denoising methods to filter redundant social signals from user-user social graph, alleviating the negative impact of such signals on recommendation performance. 

\definecolor{fig2red}{RGB}{184, 57, 57}
\definecolor{fig2blue}{RGB}{85, 102, 213}
\begin{figure}[t!]
\centering 
\includegraphics[width=0.45\textwidth]{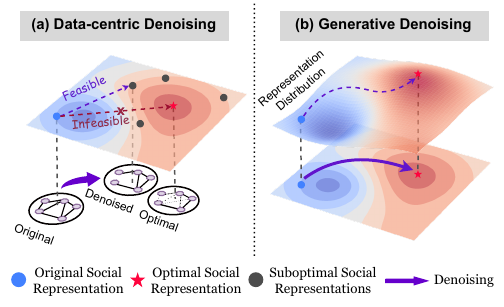}
% \vskip -0.13in
\caption{Illustration of how different social denoising methods increase the consistency between social and collaborative signals. \textcolor{fig2red}{Red} indicates high consistency, and \textcolor{fig2blue}{Blue} indicates low consistency. Existing data-centric methods (subfigure a) indirectly enhance collaborative consistency by modifying the social graph structure, but may not achieve optimal social representation due to handcrafted similarity metrics and expressive capability limitation. In contrast, the proposed generative social denoising paradigm (subfigure b) captures the underlying distribution of social graphs, directly generating optimal user representations using score-based generative diffusion models.}
%\caption{Illustration of existing social denoising methods and the proposed generative social denoising paradigm. Existing methods (subfigure a) indirectly enhance the collaborative consistency of user representations by modifying the social graph structure, which may not achieve optimal social representation due to handcrafted similarity metrics or computational constraints. In contrast, the generative social denoising paradigm (subfigure b) captures the underlying distribution of social graphs, directly generating optimal user representations using generative models. \jh{[Improvement needed - The concept of "collaborative consistency" is unclear in this figure.]}}
%The data-centric methods rely on discrete graph editing operations, constrained to one-hop local search. The SSL methods rely on simple handcrafted similar measurements to determine the positive and negative samples, resulting in imprecise optimisation objectives. In contrast, the SGM-based social denoising paradigms directly sample collaborative consistent representation from noised embeddings. \jh{[Improvement needed - Remove SSL-based methods and improve the quality of the left-hand-side subfigure]}}
\label{fig:denoise} 
\vskip -0.13in
\end{figure}

Efforts to address the low social homophily problem primarily rely on data-centric graph editing methods~\cite{jiang2024challenging, quan2023robust, tao2022revisiting, fan2023graph}, 
which denoise the social graph by deleting or rewiring edges based on human-designed heuristics, resulting in improved user representations. 
For example, GDMSR~\cite{quan2023robust} uses a transformer layer to compute user similarity with users’ interacted item lists, thereby refining the social network according to these user similarities. However, these methods face two key limitations in achieving optimal social representation consistent with collaborative signals: 
\definecolor{dark}{RGB}{77,77,77}
\newcommand{\tikzcircle}[2][red,fill=red]{\tikz[baseline=-0.55ex]\draw[#1,radius=#2] (0,0) circle ;}%
\begin{itemize}[leftmargin=*]
\item \textbf{Handcrafted Similarity Metrics}: These graph editing methods, guided by the heuristic rules, rely solely on handcrafted user similarity metrics (e.g., cosine similarity), which lacks generalizability and effectiveness in evaluating social networks. These metrics are typically limited to one-hop modifications of social relations and depend heavily on the representation capacity of the underlying backbone models in the social domain.
As illustrated in Fig.~\ref{fig:denoise} (a), handcrafted heuristics may lead to suboptimal user representations (denoted by dark circles \tikzcircle[dark, fill=dark]{0.5ex}) that are far from the optimal representation (denoted by red star \red{$\star$}).
\item \textbf{Expressive Capability Limitation}: 
These methods are unable to adaptively account for all users within a social network, which is computationally infeasible due to the vast number of users. 
The graph rewiring algorithms focus on a small, predefined subset with a fixed number of candidates, such as one-hop neighbours or users with high embedding similarity. 
This approach neglects users with similar preferences who do not conform to these predefined criteria, thereby missing opportunities to enhance the alignment between social and collaborative signals. Furthermore, extending denoising to multi-hop optimization may introduce conflicts between local and global optima in the graph editing process. 
This suggests that better social connections may not necessarily yield optimized social representations, compounded by the inherent noise in message passing approaches as validated by previous works~\cite{tao2022revisiting}.
\end{itemize}

To overcome the aforementioned limitations, we propose a fundamentally novel approach to denoise social information in recommendation systems (RS) from a \textbf{generative} perspective. 
Specifically, our key idea is to model the underlying distribution of users’ social representations and directly generate denoised social representations that maximize consistency with collaborative signals.  
This generative paradigm is highly desirable as it allows us to transcend all possible graph structure revisions and achieve the ideal social representation, which may be unreachable through graph editing steps under computational constraints. 
To instantiate this novel generative paradigm, we draw inspiration from the success of diffusion models in image restoration~\cite{luo2023image}, and propose to leverage the potential of score-based generative models (SGMs)~\cite{song2020score}, a strong yet unexplored class of diffusion models. Empowered by stochastic differential equations (SDEs), SGMs unify the diffusion processes, such as DDPM~\cite{ho2020denoising} and SMLD~\cite{song2019generative}, in continuous time steps. 
This unified continuous formulation enables a predict-and-correct generative paradigm that flexibly incorporates a wide range of numerical SDE solvers and statistical methods, thereby reducing numerical errors in the computation of diffusion models and more accurately capturing the complexities of the social denoising problem~\cite{jo2022score, fan2023generative}.

Despite the great potential, it is highly challenging to directly take advantage of SGMs for social denoising in social recommendations. 
% despite their strong potential in social denoising, directly adapting SGMs designed for image tasks to social recommendations poses significant difficulties. 
First, the optimal social network that accurately aligns with users’ preference patterns remains unknown, leading to a lack of appropriate supervision signals (i.e., training objectives) for SGMs. 
% Additionally, since SGMs were originally designed for image generation and denoising, applying them to social recommendations introduces multiple challenges, such as handling high-order graph structures and aligning signals from social and collaborative domains. 
Additionally, while SGMs were originally developed for image generation/denoising, their direct application faces two fundamental difficulties in social recommendation: (1) effectively handling high-order graph structures, and (2) aligning signals across social and collaborative domains.
To address the challenges above, we propose the \underline{S}core-based \underline{G}enerative Model for \underline{S}ocial \underline{R}ecommendation (\textbf{SGSR}) to transcend the social homophily condition in social recommendations with SGMs.
Instead of manually determining the sub-optimal embeddings from edited social graphs as supervisory signals, a curriculum learning mechanism is introduced to facilitate the modelling of social representations across social graphs with incrementally sparse levels.
Furthermore, a joint optimization method is developed to leverage collaborative signals to implicitly identify the optimal social representation. 
Our main contributions are summarized as follows:
\begin{itemize}[leftmargin=*]
    \item  We propose a score-based generative framework (\textbf{SGSR}) that effectively captures the underlying distribution of social representations and improves their collaborative consistency. 
    % \wq{To the best of our knowledge. this is the first score-based diffusion model tailored to recommendation settings.  ***we shall be more careful for this statement**}
    \item To effectively adapt the SGM in the context of social recommendation, we derive a novel classifier-free conditioning objective to achieve the personalized denoising diffusion process based on the user-item behaviour.
    \item To address training obstacles of leveraging SGMs for social denoising, we introduce an efficient training mechanism, including a curriculum learning strategy to ease the distribution learning process and a joint training mechanism to implicitly guide the denoising process. 
    \item We conduct comprehensive comparison experiments and ablation studies on three real-world datasets. The results demonstrate that the proposed SGSR framework significantly outperforms state-of-the-art baselines, validating the effectiveness of our proposed method.
\end{itemize}

\begin{comment}
\begin{itemize}[leftmargin=*]
    \item  We propose a generative framework that effectively captures the underlying distribution of social representations and improves their collaborative consistency. \wq{To the best of our knowledge. this is the first score-based diffusion model tailored to recommendation settings.  ***we shall be more careful for this statement**}
    \item To address training obstacles of leveraging SGMs for social denoising, we introduce an efficient training mechanism, including a novel conditioning objective to control the reverse diffusion process and a curriculum learning strategy to ease the learning process. \wq{*** why curriculum learning is needed?  we might briefly explain it in the introduction!!}
    \item We conduct comprehensive comparison experiments and ablation studies on three real-world datasets. The results demonstrate that the proposed SGSR framework significantly outperforms state-of-the-art baselines, validating the effectiveness of our proposed method.
\end{itemize}
\end{comment}

\section{THE PROPOSED METHOD}
\label{Method}

% In this section, we first introduce the key notations and definitions used in this paper. Subsequently, we present the basic concepts of the score-based generative diffusion models. 

In this section, we first introduce the key notations and definitions used in this paper, including the basic concepts of the score-based generative diffusion models. Subsequently, we outline the proposed framework and provide a detailed description of each model component. 

% \subsection{Preliminaries}

\subsection{Notations and Definitions} 
% \noindent \textit{\textbf{ (1) Definitions and Notations}}.
In this paper, we denote column vectors as bold lower-case letters such as $\mathbf{e}$ and matrices as bold upper-case letters such as $\mathbf{E}$. Scalar-valued functions are represented with plain letters like $g(\cdot)$, while vector-valued and matrix-valued functions are denoted with bold letters like $\mathbf{f}(\cdot)$. For embedding notations such as $\left(\mathbf{e}^{(l)}_{D}\right)_u$ and $\left(\mathbf{E}^{(l)}_{D}\right)_{\mathbf{B}}$, the inner superscript $l$ indicates the number of graph neural network layers, and the inner subscript $D$ denotes the domain of the embedding. The outer subscript $(\cdot)_u$ and $(\cdot)_{\boldsymbol{B}}$ represents the index of the embedding. 

In socially-aware recommender systems, we denote the user set as $\mathcal{U} = \{u_{1}, \dots , u_n\}$ with $n$ users and the item set as $\mathcal{V} = \{v_{1}, \dots , v_m\}$ with $m$ items. 
Given the historical user-item interaction matrix $\mathbf{R}\in\mathbb{R}^{n\times m}$, where $R_{i, j}=1$ if user $u_i$ has interacted with item $v_j$, and $R_{i, j}=0$ otherwise. Similarly, we have the social relation matrix $\mathbf{S}\in\mathbb{R}^{n\times n}$, where $S_{i, j}=1$ if user $u_i$ and user $u_j$ have an social connection, and $S_{i, j}=0$ otherwise. 

To encode collaborative and social signals into learnable embeddings, we define $\mathbf{P}_R=[\mathbf{p}_1, \dots, \mathbf{p}_n]^T\in\mathbb{R}^{n\times d}$ represent users’ embeddings in the collaborative domain, and $\mathbf{P}_S = [\mathbf{p}_1', \dots, \mathbf{p}_n']^T\in\mathbb{R}^{n\times d}$ to represent users’ embeddings in the social domain. Additionally, we define $\mathbf{Q}=[\mathbf{q}_1, \dots, \mathbf{q}_m]^T\in\mathbb{R}^{m\times d}$ to represent the item embeddings.

\subsection{An Overview of the Proposed Framework}
\label{Methodology}

\begin{figure*}[htb]
% \vskip -0.1in
\centering 
\includegraphics[width=0.95\textwidth]{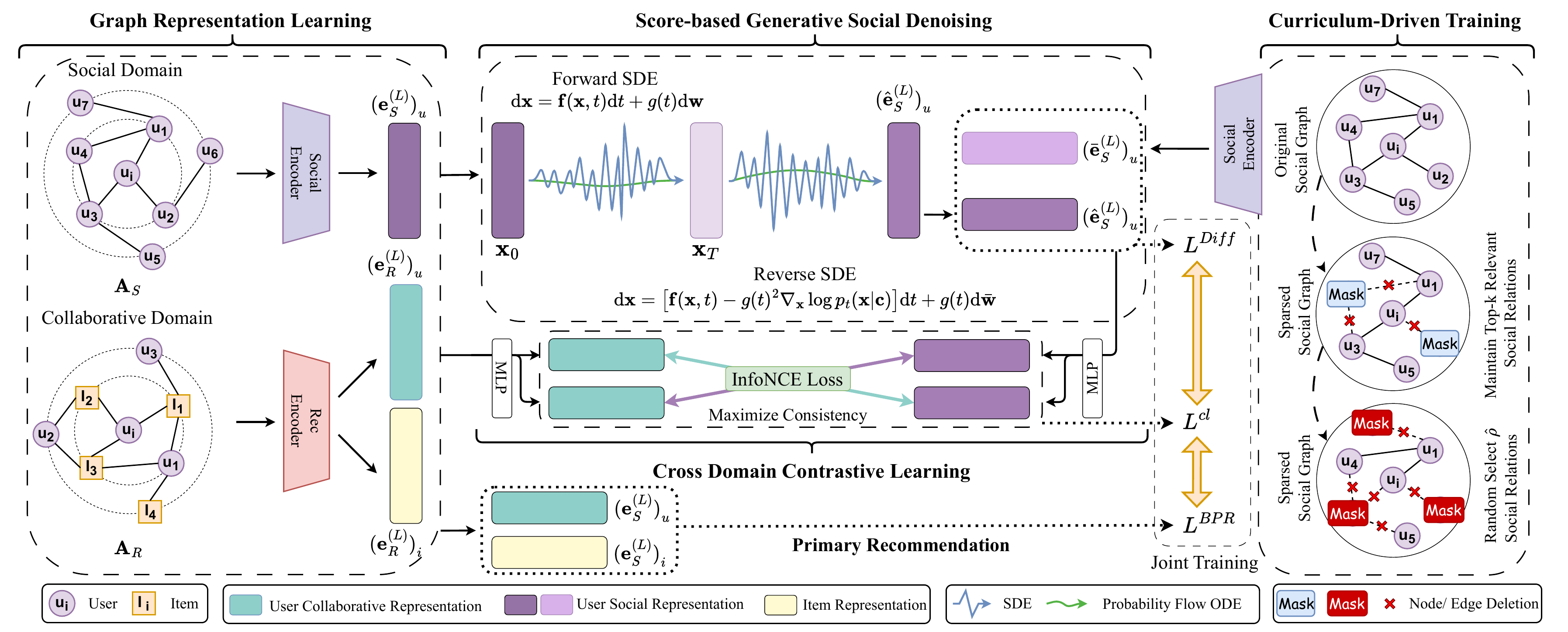}
\caption{The overall structure of the proposed SGSR. It contains four major components: graph representation learning, score-based generative social denoising, curriculum-driven training algorithm, and cross-domain contrastive learning module.}
\label{fig:framework} 
\vskip -0.2in
\end{figure*}

To alleviate the low social homophily problem in social recommendations, we propose the SGSR framework to provide accurate recommendations with generatively denoised social representations. 
Specifically, the SGSR framework incorporates a score-based social denoising model, which encodes the high-order relations in the social network into latent space and denoises the social embeddings in continuous time through a diffusion process. 
To control the social denoising model with collaborative signals, we propose a novel classifier-free training objective and enhance the training process with a curriculum learning mechanism. 
Eventually, the filtered social representations are incorporated into the collaborative domain using self-supervised learning techniques to enhance the primary recommendation task. 
The overall framework is demonstrated in Fig. ~\ref{fig:framework}.

\subsection{Score-based Generative Social Denoising} 
To address the limitations of existing graph-based social denoising methods, we adopt a controllable SDE-based diffusion process to directly optimize social representations. This approach enables the continuous refinement of social representations, guided by user interaction behavior.

% \subsubsection{\textbf{Graph Representation Learning}}
\noindent \textit{\textbf{(1) Graph Representation Learning.}} To encode the high-order social relations into the denoising framework, we first transform the original social embeddings into a latent space before the diffusion process. Specifically, we employ the widely used Graph Convolutional Network (GCN)~\cite{wu2019neural, liu2020modelling, wu2022disentangled}  as the social encoder (denoted by $\text{\textbf{SoEnc}}(\cdot)$) of our denoising framework, which follows an iterative message-passing process described as follows: 
\begin{equation}\label{eq:GCN_social}
\begin{aligned}
    \mathbf{E}_S^{(0)} &= \mathbf{P}_S, \\
    \mathbf{E}_S^{(l)} &= \phi\left(\tilde{\mathbf{A}}_S\mathbf{E}_S^{(l-1)}\mathbf{W}^{(l)}\right),\quad \forall l=1,2,\dots, L,
\end{aligned}
\end{equation}
where $L$ denotes the number of GCN layers, $\boldsymbol{W}^{(l)}$ denotes trainable transformation matrices, and $\phi$ is an activation function. The normalized adjacency matrix of the social graph is defined as $\tilde{\mathbf{A}}_S = \mathbf{D}_{\mathbf{S}}^{-1/2} \mathbf{S} \mathbf{D}_{\mathbf{S}}^{-1/2}$, where $\mathbf{D}_{\mathbf{S}}$ is the diagonal degree matrix of $\mathbf{S}$. 
%Similarly, we obtained the representations of users and items from collaborative domains, denoted as $\mathbf{P}^{(l)}_R$ and $\mathbf{Q}^{(l)}$, through arbitrary CF-based encoders $RecEnc(\cdot)$. PinSAGE~\cite{hamilton2017inductive}

Additionally, we obtain user and item representations from the collaborative domain to control the social denoising process. We adopt an arbitrary graph-based encoder (e.g., LightGCN~\cite{he2020lightgcn}, etc.) for collaborative filtering with $L$ layers, denoted by $\text{\textbf{RecEnc}}(\cdot)$, and obtain the collaborative representation as follows:
\begin{equation}\label{eq:rec_enc}
    \boldsymbol{E}_R^{(0)} = \begin{bmatrix} \boldsymbol{P}_R \\ \mathbf{Q}_R \end{bmatrix},\quad\quad \mathbf{E}_R^{(L)} = \text{\textbf{RecEnc}}\left(\mathbf{E}_R^{(0)}, \mathbf{A}_R\right),
\end{equation}
where $\mathbf{A}_R$ is the adjacency matrix of the collaborative graph. Note that the choice of $\text{\textbf{RecEnc}}(\cdot)$ is independent of the input of our generative social denoising model, indicating that the proposed SGSR is a general framework adaptable to existing recommendation backbone models.

\noindent \textit{\textbf{(2) Collaboratively Controlled Denoising Diffusion Paradigm.}} 
After capturing the long-range and non-linear relations in the social network, we apply the score-based diffusion process on the propagated social representation $\mathbf{E}_S^{(L)}$ for further denoising. Specifically, SGMs use two Stochastic Differential Equations (SDEs) to model the noising and denoising processes of the data distribution.

\noindent \textbf{Forward Process}. The forward process aims to transform the original data distribution of social representations into a prior distribution (e.g., Gaussian noise) in a continuous manner. Considering the forward diffusion process ${\{\mathbf{x}_{t}\}}_{t \in [0, T]}$ with respect to the continuous time step $t$, we first define the raw social embedding $(\mathbf{e}_S^{(L)})_{u}$ of a user $u$ as the initial state $\mathbf{x}_0$, taken from embedding matrix $\mathbf{E}_S^{(L)}$, and $\mathbf{x_{T}}$ follows a standard Gaussian prior distribution. The forward diffusion process is defined by the following SDE: 
\begin{equation}
    \mathrm{d} \mathbf{x} = \mathbf{f}(\mathbf{x}, t) \mathrm{d} t + {g}(t) \mathrm{d} \mathbf{w},
    \label{eq:SDE_forward}
\end{equation}
where $\mathbf{w}$ refers to the standard Wiener process (a.k.a. Brownian motion), and $\mathrm{d}t$ is an infinitesimal time step. The vector-valued drift coefficient $f: \mathbb{R}^{d} \rightarrow \mathbb{R}^{d}$ denotes the drift coefficient determining the data perturbation direction, and the scalar diffusion coefficient $g: \mathbb{R} \rightarrow \mathbb{R}$ represents time-dependent noise magnitude. The drift coefficient and scalar function can be manually defined to effectively guide the perturbation toward the known prior.

\noindent \textbf{Reverse Process}. The reverse-time SDE simulates the diffusion process in reverse time, reconstructing samples from a known prior that adheres to the original data distribution, which aligns with the denoising objective. We run the reverse process conditioned on the user’s collaborative information to guide the recovery of the user representation towards a collaboratively consistent direction. This can be formulated as the following conditioned reverse-time SDE~\cite{anderson1982reverse}:
\begin{equation}
    \mathrm{d}\mathbf{x} = \left\{\mathbf{f}(\mathbf{x}, t) - g(t)^{2}
    \nabla_{\mathbf{x}}  \log p_t(\mathbf{x}|\mathbf{c})\right\} \mathrm{d} t + g(t) \mathrm{d} \bar{\mathbf{w}},
    \label{eq:SDE_reverse_con}
\end{equation}
where $\bar{\mathbf{w}}$ denotes the Wiener process evolving in reverse time, $p_t$ denotes the data distribution at time $t$, and the condition vector $\mathbf{c} = (\mathbf{e}_R^{(L)})_{u}$ from $\mathbf{P}_R^{(L)}$ encodes the user’s preference pattern in the collaborative domain. The reverse SDE requires the gradient of $p_t$, known as the score $\nabla_{\mathbf{x}}  \log p_t(\mathbf{x})$, which can be estimated with a time-aware scoring neural network $\mathbf{s}_{\boldsymbol{\theta}}(\mathbf{x}_t, t, \mathbf{c})$. 

\noindent \textbf{Training}. The core of training SGMs involves optimizing the conditional scoring network $\mathbf{s}_{\boldsymbol{\theta}}(\mathbf{x}_t, t, \mathbf{c})$ using the score-matching objective, which aligns the network’s output with the actual score function: 
\begin{equation}
\hspace{-2.1mm}
\min_{\boldsymbol{\theta}}\ \mathbb{E}_{t}\left\lbrack \lambda(t) \mathbb{E}_{\mathbf{x}_{0}}\mathbb{E}_{\mathbf{x}_{t} | \mathbf{c}} \left\lVert {\mathbf{s}_{\boldsymbol{\theta}}(\mathbf{x}_t, t, \mathbf{c})\!-\!{\nabla}_{\mathbf{x}_{t}}{\log p_{t}(\mathbf{x}_{t}|\mathbf{c})}} \right\rVert^{2} \right\rbrack,
\label{eq:SDEloss_orig}
\end{equation}
where $\lambda(t)\in\mathbb{R}$ is a time-dependent weighting coefficient, $\mathbf{x}_0$ is sampled from the original data distribution $p_0$, and $\mathbf{x}_t$ is obtained from $\mathbf{x}_0$ through the forward diffusion process. 
The primary challenge in training the score model $\mathbf{s}_{\boldsymbol{\theta}}(\mathbf{x}_t, t, \mathbf{c})$ to solve this reverse-SDE lies in the analytical intractability of $\nabla_{\mathbf{x}}  \log p_t(\mathbf{x}|\mathbf{c})$.
Typically, the condition $\mathbf{c}$ is simplified to a class-conditional scenario, allowing the score to be estimated using Bayes' rule:
\begin{equation}
    \nabla_{\mathbf{x}}  \log p_t(\mathbf{x}_t|\mathbf{c}) = \nabla_{\mathbf{x}}  \log p_t(\mathbf{x}_t) + \nabla_{\mathbf{x}}  \log p_t(\mathbf{\mathbf{c}|\mathbf{x}_t}),
    \label{eq:SDE_Bayes}
\end{equation}
where $\nabla_{\mathbf{x}} \log p(\mathbf{\mathbf{c}|\mathbf{x}_t})$ is estimated by a classifier trained on the perturbed representation $\mathbf{x}_t$.
Nevertheless, training an explicit classifier for each user's social representation at every noise scale is computationally infeasible, considering that there could be thousands of users in the context of recommendations, meaning thousands of classes to train. This constraint necessitates the development of a classifier-free guidance approach for implementing score-based methods in the context of social recommendation. As an alternative, we regard the collaborative condition $\mathbf{c}$ as disentangled pseudo-labels to achieve controllable generation, thereby circumventing the need for explicit classification gradient.
Subsequently, we construct a differential loss function $\mathcal{L}^{Diff}(\mathbf{x}_0,(\mathbf{e}_R^{(L)})_{u})$ to facilitate the training of time-dependent conditional score network $s_{\theta}(\mathbf{x}_t, t, \mathbf{c})$. 
Based on Tweedie's formula and the properties of disentangled representations ~\cite{kim2022unsupervised}, 
% we mathematically prove that the $\mathcal{L}2$ minimization of conditional score matching in Eq. (\ref{eq:SDEloss_condition}) is equal to optimize the following objective (in Appendix \ref{sec:proof}):
we mathematically derive the training loss from the $\mathcal{L}2$ minimization of conditional score matching in Eq. (\ref{eq:SDEloss_orig}): 
% \jh{Due to the \texttt{\textbackslash small} command, the line spacing above is irregular. I am not sure whether this line spacing is appropriate.}
\begin{small}
\begin{equation}
\hspace{-2.7mm}
\begin{aligned}
    &\ \mathbb{E}_{t}\lbrack \lambda(t) \mathbb{E}_{\mathbf{x}_{0}}\mathbb{E}_{\mathbf{x}_{t} | \mathbf{c}} \left\lVert {\mathbf{s}_{\boldsymbol{\theta}}(\mathbf{x}_t, t, \mathbf{c})\!-\!{\nabla}_{\mathbf{x}_{t}}{\log p_{t}(\mathbf{x}_{t}| \mathbf{c})}} \right\rVert^{2} \rbrack \\
    &=\!\mathbb{E}_{t, \mathbf{x}_{0}, \mathbf{x}_{t}, \mathbf{c}}[\lambda(t){\parallel\!\mathbf{s}_{\boldsymbol{\theta}}(\mathbf{x}_t, t, \mathbf{c})\!-\!\nabla  \log p_t(\mathbf{x_t}|\mathbf{c})\!\parallel}^{2}] \\
    &=\!\mathbb{E}_{t, \mathbf{x}_{0}, \mathbf{x}_{t}, \mathbf{c}}[\lambda(t){\parallel\!\mathbf{s}_{\boldsymbol{\theta}}(x_t, t, \mathbf{c})\!-\!\nabla  \log p_t(\mathbf{x_t}, \mathbf{c})\!+\!\nabla \log p(\mathbf{c})\!\parallel}^{2}] \\
    &=\!\mathbb{E}_{t, \mathbf{x}_{0}, \mathbf{x}_{t}, \mathbf{c}} [ \lambda(t) {\parallel\!\mathbf{s}_{\boldsymbol{\theta}}(x_t, t, \mathbf{c})\!\parallel}^{2}\!-\!2\!\cdot\! \mathbf{s}_{\boldsymbol{\theta}}(x_t, t, \mathbf{c})\!\cdot\!\nabla  \log p_0t(\mathbf{x}_t|\mathbf{x}_0)]\!+\!C_1\\
    &=\!\mathbb{E}_{t, \mathbf{x}_{0}, \mathbf{x}_{t}, \mathbf{c}}[\lambda(t){\parallel\!\mathbf{s}_{\boldsymbol{\theta}}(x_t, t, \mathbf{c})\!-\!\nabla  \log p_{0t}(\mathbf{x_t}|\mathbf{x_0})\parallel}^{2}]\!-\!C_2\!+\!C_1 \\
    &=\!\mathbb{E}_{t}\lbrack \lambda(t) \mathbb{E}_{\mathbf{x}_{0}}\mathbb{E}_{\mathbf{x}_{t} | \mathbf{x}_{0},\mathbf{c}} \left\lVert {\mathbf{s}_{\boldsymbol{\theta}}(\mathbf{x}_t, t, \mathbf{c})\!-\!{\nabla}_{\mathbf{x}_{t}}{\log p_{0t}(\mathbf{x}_{t}| \mathbf{x}_{0})}} \right\rVert^{2} \rbrack\!-\!C_2\!+\!C_1,
    \label{SDE:proof}
\end{aligned}
\end{equation}
\end{small}where $p_{0t}(\mathbf{x}_{t}| \mathbf{x}_{0})$ represents the transition kernal from $\mathbf{x}_{0}$ to $\mathbf{x}_{t}$. Thus, the denoising score function ${\nabla}_{\mathbf{x}_{t}}{\log p_{t}(\mathbf{x}_{t})}$ depends only on the noising process and can be evaluated in closed form.
Finally, the SGSR employs the following classifier-free objectives to optimize the conditional scoring network $s_{\theta}(\mathbf{x}_t, t, \mathbf{c})$:
\begin{small}
\begin{equation}
\hspace{-2mm}
    \min_{\boldsymbol{\theta}}\mathbb{E}_{t}\!\left\lbrack \lambda(t) \mathbb{E}_{\mathbf{x}_{0}}\mathbb{E}_{\mathbf{x}_{t} | \mathbf{x}_{0},\mathbf{c}} \left\lVert {\mathbf{s}_{\boldsymbol{\theta}}(\mathbf{x}_t, t, \mathbf{c})\!-\!{\nabla}_{\mathbf{x}_{t}}{\log p_{0t}(\mathbf{x}_{t}| \mathbf{x}_{0})}} \right\rVert^{2} \right\rbrack\!,\!
    \label{eq:SDE_loss_dis}
\end{equation}
\end{small}where $\lambda(t)$ is empirically set to 1 based on experiments.

\noindent \textit{\textbf{SDE Instantiation.}} Despite the universality of the score-based generative paradigm, it is essential to adopt specific designs for the drift coefficient $\mathbf{f}(\boldsymbol{x}, t)$ and diffusion coefficient $g(t)$ to instantiate the SGM into practice. 
To this end, we implement the drift and diffusion coefficients using two widely adopted formulations: the \textbf{Variance Exploding (VE) SDE}, a continuous form of the Score Matching with Langevin Dynamics (SMLD)~\cite{song2019generative} model, and the \textbf{Variance Preserving (VP) SDE}, a continuous form of the well-known Denoising Diffusion Probabilistic Models~\cite{ho2020denoising}. Specifically, the forward processes of these two SDEs are formalized as follows:
\begin{align}
\mathrm{d} \mathbf{x} = 
\begin{cases}
\ \sqrt{\frac{d[{\sigma}^2(t)]}{dt}} \mathrm{d} \mathbf{w}, &(\text{VE SDE})\\
\ -\frac{1}{2}\beta(t)\mathbf{x}\ dt + \sqrt{\beta(t)}\ \mathrm{d} \mathbf{w}, &(\text{VP SDE}) 
\end{cases}\label{eq:SDE_ins}
\end{align}
where $\sigma(t)$ and $\beta(t)$ represent the predefined noise scalers. The reverse processes of the SDEs can be obtained via Eq.~\eqref{eq:SDE_reverse_con}. 

Besides, another challenge in implementing the SGM with specific SDEs lies in the computation of the denoising score function ${\nabla}_{\mathbf{x}_{t}}{\log p_{t}(\mathbf{x}_{t} | \mathbf{x}_{0} )}$, which is an indispensable part of the reverse process. Fortunately, the specifications of the drift and diffusion coefficients enable a closed-form solution for the Gaussian transition kernel $p_{t}(\mathbf{x_t}|\mathbf{x_0})$, facilitating an efficient training process.
\begin{small}
\begin{align}
p_{t}(\mathbf{x}_t|\mathbf{x}_0)\!=\!
\begin{cases}
\!\mathcal{N}\left(\mathbf{x}_t; \mathbf{x}_0,[{\sigma}^2(t)-{\sigma}^2(0) ]\mathbf{I}\right),\!&(\text{VE SDE})\\
\!\mathcal{N} (\mathbf{x}_t; \mathbf{x}_{0}{e}^{-\frac{1}{2}\int^{t}_{0} \beta(s)ds}, \mathbf{I}-\mathbf{I}{e}^{-\int^{t}_{0} \beta(s)ds} ),\!&(\text{VP SDE}) 
\end{cases}
\end{align}
\end{small}
% Thus, with the closed-form Gaussian transition kernel, it is straightforward to derive the corresponding probability density function and obtain the closed-form solution of the score functions. 
The closed-form Gaussian transition kernel enables the direct derivation of both the probability density function and the analytical solutions of the score functions.

\noindent \textit{\textbf{Social Reperesentation Denoising.}}\label{sec:sampling_process}
After training the score-based social denoising model, we obtain the scoring network $\mathbf{s}_{\boldsymbol{\theta}}$ and use it to denoise users’ social representations. 
Given a user's raw social representation $(\mathbf{e}_S^{(L)})_u$, the model first perturbs the raw social representation in a continuous manner using either the forward VP SDE or VE SDE in Eq.~\eqref{eq:SDE_ins} to obtain the output $\mathbf{x}_T$. 
It then reconstructs the denoised representation $(\mathbf{\hat{e}}_S^{(L)})_u$ by iteratively solving the collaboratively conditioned reverse-SDE (Eq.~\ref{eq:SDE_reverse_con}) with off-the-shelf SDE initial-value problem (IVP) numerical solvers~\cite{platen2010numerical}.
To obtain the numerical solution of a given SDE, we adopt the Predictor-Corrector (PC) method~\cite{song2020score}, a flexible and powerful framework for solving SDEs inherent in SGMs. 
The iterative steps of the Predictor-Corrector solver first use the prediction of numerical SDE solvers like the Euler–Maruyama method, and then correct the predicted trajectory with score-based MCMC methods, such as Langevin MCMC, achieving a balance of numerical accuracy and efficiency. 
The Algorithm ~\ref{alg:sampling} outlines the inference process of the denoising model, incorporating annealed Langevin dynamics as a corrector ($\epsilon_i$ represents the step size for Langevin Dynamics).
% The Algorithm 1 demonstrates the details of the inference process of the denoising model cooperated with the annealed Langevin dynamics as a corrector.the Euler–Maruyama method~\cite{mil1975approximate}Langevin MCMC~\cite{grenander1994representations}

\renewcommand{\thealgorithm}{1}
\begin{algorithm}[t]
\caption{Inference Process of SGM} 
\begin{algorithmic}
    \Require 
    $\mathcal{M}$: Discretized diffusion steps, $\mathcal{N}$: Corrector steps
    \State Initialize $\mathbf{x}_{\mathcal{M}} \sim p_{t}(\mathbf{x}_t|\mathbf{x}_0) $
    \For{$i = \mathcal{M}-1$ to $0$} 
    \State \textbf{Option I: VE SDE} (predictor):
        \State $\quad\bar{\mathbf{x}}_i \leftarrow \mathbf{x}_{i+1} + ({\sigma}^2_{i+1}- {\sigma}^2_{i}) \mathbf{s}_{\boldsymbol{\theta}}(\mathbf{x}_{i+1}, \sigma_{i+1}) $
        \State $\quad\mathbf{z} \sim \mathcal{N}(0, \mathbf{I})$
        \State  $\quad\mathbf{x}_i \leftarrow \bar{\mathbf{x}}_i + \sqrt{{\sigma}^2_{i+1}- {\sigma}^2_{i}} \mathbf{z}$
    \State \textbf{Option II: VP SDE} (predictor):
        \State $\quad\bar{\mathbf{x}} \leftarrow (2-\sqrt{1-\beta_{i+1}}){x}_{i+1}+\beta_{i+1} \mathbf{s}_{\boldsymbol{\theta}} (\mathbf{x}_{i+1}, i+1) $
        \State $\quad\mathbf{z} \sim \mathcal{N}(0, \mathbf{I})$
        \State  $\quad\mathbf{x}_i \leftarrow \bar{\mathbf{x}}_i +\sqrt{\beta_{i+1}} \mathbf{z}$
        \For {$j = 1$ to $\mathcal{N}-1$} 
            \State $\mathbf{z} \sim \mathcal{N}(0, \mathbf{I})$
            \State \textbf{Option I: VE SDE} (corrector):
                \State $\quad\mathbf{x}_i \leftarrow \mathbf{x}_i + \epsilon_i \mathbf{s}_{\boldsymbol{\theta}} (\mathbf{x}_{i}, \sigma_{i}) + \sqrt{2\epsilon_i} \mathbf{z}$
            \State \textbf{Option II: VP SDE} (corrector):
                \State $\quad\mathbf{x}_i \leftarrow \mathbf{x}_i + \epsilon_i \mathbf{s}_{\boldsymbol{\theta}} (\mathbf{x}_{i}, i) + \sqrt{2\epsilon_i} \mathbf{z}$
        \EndFor
    \EndFor
    \Ensure
    $x_0$: Denoised social representation
\end{algorithmic}
\label{alg:sampling}
% \vskip -0.1in
\end{algorithm}

\subsection{Curriculum-Driven Training of the Social Denoising Model} 

Since the optimal social representation that maximizes consistency with collaborative signals is difficult to determine, training the SGM for social graph denoising presents significant challenges due to the lack of ground truth and supervision signals. For example, in image restoration tasks, SGMs use deteriorated images as input and compute the loss function between the ground-truth image and the denoised image. This straightforward training strategy is inapplicable in the context of social denoising, as the ground-truth optimal social representation is not accessible.
To overcome this limitation, we employ a curriculum learning strategy that enables the SGM to perceive social representations across social graphs with varying levels of sparsity. This approach potentially encompasses the optimal social representations, which the SGM model can identify automatically with collaborative signals.

The two-stage curriculum training mechanism implements a progressive learning paradigm, gradually exposing the SGM to increasingly sparse social distributions. The sparser social distributions, characterized by fewer relationships, encapsulate reduced learnable information, posing greater challenges for models to learn the distributions ~\cite{wang2021survey}. This method forms an easy-to-hard training strategy, enhancing the robustness of the learning process. Specifically:
\begin{itemize}[leftmargin=*]
\item In the initial stage, the model retains the percentage $\rho$ of social connections that exhibit the most similar preference patterns for $N$ epochs. Given the collaborative representations of user $i$ and $j$ from $RecEnc(\cdot)$ as $(\mathbf{e}_R^{(L)})_{u_i}$ and $(\mathbf{e}_R^{(L)})_{u_j}$, we leverage the inner product to measure the similarity $\mathrm{sim} = (\mathbf{e}_R^{(L)})_{u_i} \cdot {(\mathbf{e}_R^{(L)})^{T}_{u_j}} $. 
% \jh{For $sim, MLP, \mathcal{L}^{Diff}, \mathcal{L}^{cl}, \mathcal{L}^{BPR}$, how about using \texttt{\textbackslash mathrm} or \texttt{\textbackslash text}? I mean $\mathrm{sim}, \mathrm{MLP},\mathcal{L}^\mathrm{Diff}, \mathcal{L}^{\mathrm{CL}}, \mathcal{L}^{\mathrm{BPR}}$. This will make the notation of these functions consistent with some pre-defined functions like $\exp, \cos, \log$. }
\item In the second stage, a random selection mechanism keeps $\hat{\rho}$ percentage of social relations for the subsequent $M$ epochs in the curriculum period, facilitating SGM's access to a broader range of social representations.
\end{itemize}

This approach encourages the SGM to model the relationship between collaborative conditions and social patterns across varying levels of graph sparsity. The initial process focuses on user-level sparsification, selectively removing seemingly irrelevant social connections that are more readily learnable ~\cite{zhao2021heterogeneous}, while the latter method implements a holistic network sparsification, challenging the SGM to model more diverse representations ~\cite{ma2018randomly}.

The training objectives in Eq. (\ref{eq:SDE_loss_dis}) can be updated with the sparse social representation $\bar{\mathbf{E}}_S^{(L)}$ obtained by the same $SocEnc(\cdot)$ from the post-edition social graph. The updated loss, $\mathcal{L}^{Diff}$, is formulated as:  
% \jh{For $p_{0t}$, how about using $p_{0, t}$?}
\begin{small}
\begin{equation}
\hspace{-2mm}
    \min_{\boldsymbol{\theta}} \mathbb{E}_{t}\left\lbrack \lambda(t) \mathbb{E}_{\mathbf{\bar{x}}_{0}}\mathbb{E}_{\mathbf{\bar{x}}_{t} | \mathbf{\bar{x}}_{0},\mathbf{c}} \left\lVert {\mathbf{s}_{\boldsymbol{\theta}}(\mathbf{x}_t, t, \mathbf{c})\!-\!{\nabla}_{\mathbf{x}_{t}}{\log p_{0t}(\mathbf{\bar{x}}_{t}| \mathbf{\bar{x}}_{0})}} \right\rVert^{2} \right\rbrack,
    \label{eq:diff_loss}
\end{equation}
\end{small}
\noindent where $\bar{\mathbf{x}}_0 = (\mathbf{\bar{e}}_S^{(L)})_u \sim \bar{\mathbf{E}}_S^{(L)}$ denotes the social embedding of sampled user $u$, $\mathbf{\bar{x}_t} \sim p_{0t} (\mathbf{\bar{x}_t}|\mathbf{\bar{x}_0})$ represents the distribution of data at the intermediate diffusion step $t$, $\mathbf{c}$ representes the corresponding collaborative representation as disentangled condition.
Note that the SGSR does not designate a specific sparsity level as the optimal representation; instead, it is determined by the recommendation signal through joint training strategy as described in ~\ref{sec:joint_training}.

\renewcommand{\thealgorithm}{2}
\begin{algorithm}[thp]
\caption{Training Process of SGSR} 
\begin{algorithmic}
    \Require social relation matrix $\mathbf{S}$, user-item interaction matrix $\mathbf{R}$, and randomly initialized parameters $\boldsymbol{\theta}$
    % \For{ epoch = 0 to E} 
    % \State epoch \leftarrow 0
    \Repeat
    \If {epoch == 0 or epoch $\% (M+N)$ == 0}
        \State Update the sparsed social matrix $\mathbf{\bar{S}}$ by retaining 
        \State the top-$k$ social relations 
    \ElsIf {epoch $\% N$ == 0:}
        \State Update the sparsed social matrix $\mathbf{\bar{S}}$ by randomly
        \State selecting $\hat{\rho}$ percentage social relations
    \EndIf
    \State Sample a batch of users $u$, each with a positive item 
    \State and a negative sample
    \For{each batch of users} 
        \State Obtain the representations of users and items from 
        \State social domain $\mathbf{E}_S^{(L)}$ and collaborative domain $\mathbf{E}_R^{(L)}$
        \State Sample $t \sim \mathbf{\mathcal{U}}(1, T)$
        \State $\mathbf{x}_0 \leftarrow (\mathbf{e}_S^{(L)})_u \sim \mathbf{E}_S^{(L)}$
        \State Compute $\mathbf{x}_t$ from $\mathbf{x}_0$ via Eq. (\ref{eq:SDE_ins})
        \State Compute the $\mathcal{L}^{Diff}$ by Eq. (\ref{eq:diff_loss})
        \State Sample denoised social representation $\left(\hat{\mathbf{e}}_S^{(L)}\right)_u$ via 
        \State Algorithm \ref{alg:sampling}
        \State Compute the $\mathcal{L}^{cl}$ by Eq. (\ref{eq:NCEloss})
        \State Compute the $\mathcal{L}^{BPR}$ by Eq. (\ref{eq:BPRloss})
        \State Take gradient descent step to update $\boldsymbol{\theta}$
    \EndFor
    \Until {converged}
    \Ensure
    $\boldsymbol{\theta}$: Optimized parameters
\end{algorithmic}
\label{alg:training}
\end{algorithm}

\subsection{Score-based Denoising Model Optimization}
Besides explicit conditioning on the observed user behaviour patterns, the SGSR leverages the recommendation signal to implicitly guide the SGM in identifying the optimal social representation. On the other hand, to integrate the produced social representation into the collaborative domain, the SGSR introduces an auxiliary SSL loss to augment the primary recommendation task.

\noindent \textit{\textbf{(1) Cross-domain Contrastive Learning.}}
The stochastic Wi-\\ener process and exposure bias in the interactive sampling process may introduce undesired noise to the generated social representation~\cite{ning2024elucidating}. To mitigate these biases and effectively leverage the learned social representation distribution, the SGSR incorporates contrastive learning techniques, known for their high efficiency in knowledge transfer ~\cite{yu2021self, wu2022disentangled}. The SGSR enhances the RS with social information by maximizing the consistency of user representations across the social and collaborative domains.

As described in \ref{sec:sampling_process}, we construct the denoised social representation  $\hat{\textbf{E}}_S^{(L)}$, produced by the SDE-based diffusion process as the self-supervised signal. To integrate social information into the collaborative domain, we adopt the InfoNCE loss ~\cite{oord2018representation} as our contrastive learning objective to optimize the user representations by maximizing the agreement between  $\hat{\textbf{E}}_S^{(L)}$ and $\mathbf{P}^{(L)}_R$, which can be defined as:
\begin{equation}
    \mathcal{L}^{cl} = \sum_{u \in \mathcal{U}} -\log \frac{\varsigma(\mathbf{\hat{e}}_u, \mathbf{p}_{u})}{\varsigma(\mathbf{\hat{e}}_u, \mathbf{p}_{u}) + \sum\limits_{v \in \mathcal{U}, v \neq u} \varsigma(\mathbf{\hat{e}}_u, \mathbf{p}_{v})) }, 
    \label{eq:NCEloss}
\end{equation}
where $\mathbf{\hat{e}}_u = \mathrm{MLP}\left((\mathbf{\hat{e}}_S^{(L)})_u\right)$, $ \mathbf{p}_{u} = \mathrm{MLP}\left((\mathbf{e}_R^{(L)})_{u}\right)$, and $ \mathbf{p}_{v} = \mathrm{MLP}\left((\mathbf{e}_R^{(L)})_{v}\right)$ denotes the projected social representation of user $u$ and collaborative representation of user $u$ and $v$, respectively.These representations are mapped into a shared semantic space using a multi-layer perception $\mathrm{MLP}(\cdot)$. The similarity function $\varsigma(\mathbf{\hat{e}}_{u}, \mathbf{p}_{*}) = \exp\left({\cos(\mathbf{\hat{e}}_{u}, \mathbf{p}_{*})} / \tau \right)$ measures the representation similarity based on cosine similarity, where $\tau$ is the temperature parameter.

% where $\mathbf{\hat{e}}_u = MLP\left((\mathbf{\hat{e}}_S^{(L)})_u\right)$ and $ \mathbf{p}_{u} = MLP\left((\mathbf{e}_R^{(L)})_{u}\right)$ denotes the social representation and collaborative representation of user $u$ projected to the same semantic space by multi-layer perceptions $MLP(\cdot)$, $\varsigma(\mathbf{\hat{e}}_{u}, \mathbf{p}_{u}) = \exp\left({\cos(\mathbf{\hat{e}}_{u}, \mathbf{p}_{u})} / \tau \right)$, and $\tau$ is the temperature parameter. 

\noindent \textit{\textbf{(2) Joint Training Mechanism.}}
\label{sec:joint_training}
The SGSR leverages the recommendation signal to guide the social representation refinement trajectory with the well-established Bayesian Personalized Ranking (BPR) loss, denoted as $\mathcal{L}^{BPR}$ ~\cite{rendle2012bpr}.
\begin{equation}
    \mathcal{L}^{BPR} = \sum_{i\in\mathcal{V}_+, j\notin\mathcal{V}_+} -\log \sigma (\hat{p}_{ui} - \hat{p}_{uj}),
\label{eq:BPRloss}
\end{equation}
where $\mathcal{V}_+$ represents the set of items interacted by user $u$, $\sigma(\cdot)$ denotes the sigmoid function, $\hat{p}_{ui}$ and $\hat{p}_{uj}$ represent the predicted scores for user $u$ over the positive item and negative sample, respectively. The joint optimization loss for the SGSR is:
\begin{equation}
    \mathcal{L}= \mathcal{L}^{Diff} + \lambda_1 \mathcal{L}^{BPR} + \lambda_2 \mathcal{L}^{cl},
    \label{loss}
\end{equation}
where $\lambda_1$ and $\lambda_2$ are hyperparameters to balance the social denoising score matching and the recommendation accuracy.
Details of the training process of the denoising model are shown in the Algorithm ~\ref{alg:training}.

\subsection{Complexity Analysis of SGSR}

This section analyzes the space and time complexity of the proposed SGSR framework.

\noindent \textit{\textbf{(1) Space Complexity Analysis.}} 
% \jh{For these complexity analysis, how about moving them to the ``Score-based Denoising Model Optimization'' Section?}
The space complexity analysis of SGSR focuses on two primary components: graph encoders and the reverse diffusion process. The social encoder $\text{\textbf{SoEnc}}(\cdot)$ requires $\mathcal{O}(|\mathbf{S}|+ d^2 + |U|d)$, assuming the input and output dimensions of GCN are equal, where $|\mathbf{S}|$ represents the number of social connections. Meanwhile, the collaborative encoder $\text{\textbf{RecEnc}}(\cdot)$ (e.g., LightGCN) needs $\mathcal{O}(|\mathbf{R}| + (|\mathcal{U}|+|\mathcal{V}|)d)$, with $|\mathbf{R}|$ representing the number of item interactions. 
Aditionally, the reverse diffusion process and the scoring network $\mathbf{s}_{\boldsymbol{\theta}}$ jointly require 
% $\mathcal{O}(dT + \sum_{\ell=1}^{\mathcal{L}}{d_{\ell-1} \cdot d_{\ell}})$ 
$\mathcal{O}(dT + {L} \cdot d^2)$ 
space, where $T$ is the number of diffusion step, ${L}$ is the number of layers in the scoring network $\mathbf{s}_{\boldsymbol{\theta}}$, and $d$ represents the dimensionality of each layer, assumed to be consistent across layers for brevity.

% \jh{How about assuming that all the dimensions $d_1, \ldots,d_\mathcal{L}$ equals to $d$, or they are just at the same order as $d$? This can simplify the results and make it clearer. BTW, for the notation of number of layers, do we use $\mathcal{L}$ or $L$? We have already used $\mathcal{L}$ for loss functions.}

\noindent \textit{\textbf{(2) Time Complexity Analysis.}} 
From a theoretical perspective, the time complexity analysis of SGSR primarily consists of two components: GNN-based representation aggregation and diffusion-based denoising process. Initially, SGSR utilizes GCN to produce social embeddings. Each layer involves message passing and feature aggregation operations, yielding a time complexity of $\mathcal{O}(|\mathbf{S}|d + |\mathcal{U}|d^2)$.
% where $|\mathbf{S}|$ represents the number of social connections. 
The time complexity of $\text{\textbf{RecEnc}}(\cdot)$ depends on the specific GNN encoder. Taking LightGCN as an example, it requires $\mathcal{O}(|\mathbf{R}|d + (|\mathcal{U}|+|\mathcal{V}|)d)$.
% where $|\mathbf{R}|$ represents the number of item interactions.
Subsequently, for each batch, the forward diffusion process requires $\mathcal{O}(Bd)$ calculations to perturb the social representation, while the one-step inference costs $\mathcal{O}(B\cdot {L} \cdot d^2)$, where $B$ denotes the batch size.

\section{Experiment}
\label{Experiment}

In this section, we evaluate the effectiveness of the proposed SGSR framework through extensive experiments.  

\subsection{Experiment Settings}

\noindent  \textit{\textbf{ (1) Datasets.}}
The experiments are conducted on three real-world social recommendation datasets: Ciao~\footnote{Ciao and Epinions datasets: \url{https://www.cse.msu.edu/~tangjili/trust.html}}, Epinions$^1$, and Dianping~\footnote{\url{https://lihui.info/data/dianping/}}. 
Each dataset records users’ ratings on items on a scale of $[1, 5]$ and their social connections.  
These three datasets have been shown to satisfy the low social homophily assumption (with graph-wise homophily ratio smaller than 0.1 ~\cite{jiang2024challenging}) and used to evaluate the social denoise task according to the previous literature ~\cite{tao2022revisiting}.
Following a widely used setting~\cite{fan2019graph, yu2021self, jiang2024challenging, tao2022revisiting}, we transform the explicit ratings into implicit interactions by only keeping interactions with ratings above 3, and then filter out users and items with fewer than three interactions to ensure dataset quality. For all three datasets, we split the data into training, validation, and testing sets in a ratio of 8:1:1. The statistics of the datasets are summarized in Table~\ref{tab:dataset_statistics}.

\begin{table}[thp]
\centering
% \vskip -0.15in
\caption{Statistics of the datasets}
% \vskip -0.15in
\begin{tabular}{c|ccc}
\toprule
\toprule 
\textbf{Dataset} & \textbf{Ciao} & \textbf{Epinions} & \textbf{Dianping}\\
\hline
\#Users & 7,317 & 18,088 & 147,918  \\
% \hline
\#Items & 104,975 & 261,649 & 11,123 \\
% \hline
\#Interactions & 283,319 & 764,352  & 2,149,675 \\
Interaction Density & 0.0368\%  & 0.0161\% & 0.1306\% \\
\hline
\#Relations & 111,781  & 355,813 & 629,618 \\
% \hline
Relation Density & 0.2087\%  & 0.1087\% & 0.0028\% \\
\hline
Graph-wise Homophily Rate & 0.0071  & 0.0015 & 0.0089 \\
\bottomrule
\bottomrule

\end{tabular}
\label{tab:dataset_statistics}
\vskip -0.15in
\end{table}

\noindent  \textit{\textbf{ (2) Evaluation Metrics.}} 
To evaluate top-$K$ recommendations from implicit user feedback, we use two standard ranking metrics: Recall@$K$ (R@$K$) and Normalized Discounted Cumulative Gain (N@$K$). We set $K\in\{5, 10\}$ for evaluation and rank items among all candidates to conduct a full-rank evaluation protocol. To ensure statistical reliability, all experiments were repeated five times with results reported as mean values.

\begin{table*}[t]
\caption{Comparison of overall performance on three social recommendation datasets. The best performance is highlighted in bold, and the second best is underlined. \%Improve denotes the relative improvement of SGSR over the strongest baseline.}
% \vskip -0.1in
\centering
\begin{tabular}{c|cccc|cccc|cccc}
\toprule
\toprule
\textbf{Dataset}  & \multicolumn{4}{c|}{\textbf{Ciao}} & \multicolumn{4}{c}{\textbf{Epinions}} & \multicolumn{4}{|c}{\textbf{Dianping}} \\
\cmidrule{1-13}
\textbf{Method}   & \textbf{R@10} & \textbf{R@5} & \textbf{N@10} & \textbf{N@5} & \textbf{R@10} & \textbf{R@5} & \textbf{N@10} & \textbf{N@5} & \textbf{R@10} & \textbf{R@5} & \textbf{N@10} & \textbf{N@5} \\
% \hline
\hline
DiffRec & 0.0564 & 0.034 & 0.0372 & 0.0268 & 0.0368 & 0.0224 & 0.0246 & 0.0197 & 0.0519 & 0.0304 & 0.0337 & 0.0269 \\
DreamRec & 0.0552 & 0.0349 & 0.0352 & 0.0267 & 0.0311 & 0.0219 & 0.0238 & 0.0201 & 0.0517 & 0.0307 & 0.0336 & 0.0271 \\
DreamRec+ & 0.0556 & 0.0346 & 0.0351 & 0.0259 & 0.0319 &0.0223 & 0.0241 & 0.0198 & 0.0527 & 0.0314 & 0.0342 &  0.0272\\
BSPM & 0.0559 & 0.0326 & 0.0365 & 0.0282 & 0.0402 & 0.0250 & 0.0254 & 0.0211 & 0.0616 & 0.0365 & 0.0407 & 0.0319 \\
GiffCF & 0.0579 & 0.0379 & 0.0397 & 0.0323 & 0.0417 & 0.0268 & 0.0280 & 0.0223 & 0.0655 & 0.0424 & 0.0428 & 0.0338\\
\hline
SocialMF & 0.0331 & 0.0174 & 0.0247 & 0.0132 & 0.0243 & 0.0154 & 0.0148 & 0.0112 & 0.0330 & 0.0198& 0.0184 & 0.0154\\
NGCF &  0.0559 & 0.0337 & 0.0361 & 0.0283 & 0.0362 & 0.0235 & 0.0249 & 0.0217 & 0.0523 & 0.0312 & 0.0346 & 0.0273 \\
LightGCN & 0.0579 & 0.0351 & 0.0392 & 0.0306 & 0.0428 & 0.0273 & 0.0295 & 0.0241 & 0.0526 & 0.0314 & 0.0361 & 0.0288 \\
DiffNet  &  0.0461 & 0.0286 & 0.0303 & 0.0238 & 0.0366 & 0.0225 & 0.0241 & 0.0192 & 0.0538 & 0.0317 & 0.0349 & 0.0271 \\
SEPT &  0.0599 & 0.0368 & 0.0403 & 0.0322 & 0.0445 & 0.0275 & 0.0291 & 0.0232 & 0.0671 & 0.0405 & 0.0451 & 0.0362 \\
MHCN &  0.0627 & 0.0383 & 0.0412 & 0.0326 & \underline{0.0456} & 0.0282 & 0.0303 & 0.0239 & 0.0699 & \underline{0.0423} & 0.0463 & \underline{0.0379}\\ 
\hline
DESIGN & 0.0623 & \underline{0.0388} & 0.0411 & 0.0327 & 0.0451 & \underline{0.0285} & 0.0301 & \underline{0.0241} & \underline{0.0701} & 0.0421 & 0.0458 & 0.0365 \\
DSL & 0.0603 & 0.0353 & \underline{0.0414} & 0.0324 & 0.0448 & 0.0260 & 0.0297 & 0.0235 & 0.0688& 0.0403 & 0.0415 & 0.0358 \\
GBSR & 0.0585 & 0.0386 & 0.0405 & 0.0301 & 0.0443 & 0.0279 & 0.0288 & 0.0238 & 0.0699 & 0.0420 & \underline{0.0464} & 0.0373\\
SHaRe & 0.0617 & 0.0376 & 0.0407 & \underline{0.0328} & 0.0454 & 0.0283 & \underline{0.0304} & 0.0241 & 0.0687 &0.0419 & 0.0453 & 0.0362\\
EIISRS & 0.0601 & 0.0372 & 0.0402 & 0.0322 & 0.0441 & 0.0277 & 0.0296 & 0.0239 & 0.0683 & 0.0407 & 0.0462 & 0.0360 \\
RecDiff & 0.0604 & 0.0380 & 0.0404 & 0.0325  & 0.0445 & 0.0272 & 0.0293 & 0.0231 & 0.0689 & 0.0414 & 0.0463 & 0.0368 \\
GDSSL & \underline{0.0627} & 0.0377 & 0.0408 & 0.0317 & 0.0431 & 0.0265 & 0.0277 & 0.0220 & 0.0676 & 0.0417 & 0.0456 & 0.0355 \\
\hline
SGSR (Ours) &  \textbf{0.0645} & \textbf{0.0405} & \textbf{0.0425} & \textbf{0.0346} & \textbf{0.0470} & \textbf{0.0297} & \textbf{0.0315} & \textbf{0.0246 } & \textbf{0.0714} & \textbf{0.0439}& \textbf{0.0482} & \textbf{0.0390} \\
\%Improve & 2.87\% & 4.38\% & 2.66\% & 5.49\% & 3.07\% & 4.21\% & 3.62\% & 2.07\% & 1.85\% & 3.78\% & 3.88\% & 2.90\%  \\
\bottomrule
\bottomrule
\end{tabular}

\label{OverallPerfromance}
\end{table*}
\begin{table}[!ht]

\caption{Comparison of performance of social recommendation on Epinions datasets. The best performance is highlighted in bold, and the second best is underlined. \%Improve denotes the relative improvement of SGSR over the strongest baseline.}
\centering
% \resizebox{0.9\linewidth}{!}{
% \setlength{\tabcolsep}{1.5mm}{
\begin{tabular}{c|cccccc}
\toprule
\toprule
\textbf{Dataset}  & \multicolumn{6}{c}{\textbf{Epinions}} \\
\cmidrule{1-7}
% \textbf{Method}   & \textbf{R@40} & \textbf{R@20} & \textbf{R@3} & \textbf{N@40} & \textbf{N@20} & \textbf{N@3} \\
\textbf{Method}   & \textbf{R@40} & \textbf{R@20} & \textbf{R@3} & \textbf{N@40} & \textbf{N@20} & \textbf{N@3} \\
\hline
DiffNet & 0.0906 & 0.0624 & 0.0169 & 0.0394 & 0.0327 & 0.0175  \\
SEPT    & 0.0919 & 0.0657 & 0.0171 & 0.0417 & 0.0349 & 0.0202  \\
MHCN    & 0.0939 & 0.0666 & 0.0178 & 0.0425 & 0.0352 & \underline{0.0203}  \\ 
\hline
DESIGN  & 0.0935 & 0.0659 & 0.0175 & \underline{0.0432} & 0.0353 & 0.0201  \\
DSL     & 0.0926 & 0.0640 & 0.0153 & 0.0419 & 0.0347 & 0.0189  \\
GBSR    & 0.0913 & 0.0656 & 0.0167 & 0.0427 & 0.0351 & 0.0197  \\
SHaRe   & 0.0934 & \underline{0.0673} & 0.0184 & 0.0430 & 0.0356 & 0.0195  \\
EIISRS  & \underline{0.0944} & 0.0661 & 0.0181 & 0.0431 & 0.0355 & 0.0198  \\
RecDiff & 0.0929 & 0.0669 & \underline{0.0187} & 0.0426 & \underline{0.0357} & 0.0200  \\
GDSSL   & 0.0941 & 0.0671 & 0.0183 & 0.0429 & 0.0342 & 0.0196  \\
\hline
SGSR & \textbf{0.1013} & \textbf{0.0697}& \textbf{0.0205} & \textbf{0.0450} & \textbf{0.0372}& \textbf{0.0214} C\\
\%Improve & 7.31\% & 3.57\% & 9.63\% & 4.17\% & 4.20\% & 5.42\% \\
\bottomrule
\bottomrule
\end{tabular} %} }
\label{Perfromance}
\vskip -0.15in
\end{table}

\noindent  \textit{\textbf{ (3) Baselines.}} 
We evaluate the SGSR's performance against the state-of-the-art baselines across several categories: vanilla social recommendation models (SocialMF~\cite{jamali2010matrix}, NGCF~\cite{wang2019neural}, LightGCN~\cite{he2020lightgcn}, DiffNet~\cite{wu2019neural}, SEPT\cite{yu2021socially}, MHCN~\cite{yu2021self}), diffusion recommendation models (DiffRec~\cite{wang2023diffusion}, DreamRec~\cite{yang2024generate}, DreamRec+, BSPM~\cite{choi2023bspm}, GiffCF~\cite{zhu2024graph}), and social graph denoising methods (DESIGN~\cite{tao2022revisiting}, DSL~\cite{wang2023denoised}, GBSR~\cite{yang2024graph}, SHaRe~\cite{jiang2024challenging}, EIISRS~\cite{chen2024social}, RecDiff~\cite{li2024recdiff}, GDSSL~\cite{li2024graph}). Descriptions of these baseline methods are provided as follows:
% Descriptions and implementation details of these baseline methods are provided in Appendix~\ref{sec:appendix_baseline}.

\begin{itemize}[leftmargin=*]
\item DiffRec ~\cite{wang2023diffusion}: Diffrec initially introduces the diffusion algorithm to the recommendation context, which perturbs the user's historical interaction over the whole item vector.
\item DreamRec ~\cite{yang2024generate}: DreamRec accomplishes the sequence recommendation by generating the consequent item embedding and recommending with item retrieval techniques.
\item DreamRec+: DreamRec+ combines the social information with the generation guidance embedding to adopt the DreamRec to social recommendation.
\item BSPM ~\cite{choi2023bspm}: BSPM applies score-based generative models to collaborative filtering by introducing a blurring and sharpening process on the interaction matrix.
\item GiffCF ~\cite{zhu2024graph}: GiffCF applies the diffusion algorithm utilizing the heat equation to the item-item similarity graph for collaborative filtering with implicit feedback.
\item SocialMF ~\cite{jamali2010matrix}: SocialMF propagates the social information into the matrix factorization model.
\item NGCF ~\cite{wang2019neural}: NGCF leverages the graph convolutional network to capture collaborative signals by propagating embeddings on the user-item interaction graph.
\item LightGCN ~\cite{he2020lightgcn}: LightGCN simplifies NGCF by removing the feature transformation and nonlinear activation layers, specifically designed for CF tasks.
\item DiffNet ~\cite{wu2019neural}: DiffNet implements layer-wise propagation mechanisms to model the diffusion of social influence in recommendation systems.
\item MHCN ~\cite{yu2021self}: MHCN incorporates multi-channel hypergraph convolution operations to capture complex high-order correlations among users, items, and social relations.
\item SEPT ~\cite{yu2021socially}: SEPT introduces a socially aware self-supervised learning (SSL) framework that incorporates tri-training to capture supervisory signals from both the node itself and its neighbouring nodes.
\item DESIGN ~\cite{tao2022revisiting}: DESIGN leverages knowledge distillation to integrate information from both the user-item interaction graph and the social graph. 
\item DSL~\cite{wang2023denoised}: DSL leverages SSL techniques to denoise the social representation by aligning the embedding with those who share similar preference patterns.
\item GBSR~\cite{yang2024graph}): GBSR introduces a theoretically-inspired framework to maximize the mutual information between the denoised social graph and the interaction matrix.
\item SHaRe~\cite{jiang2024challenging}: SHaRe designs the graph rewiring technique to discretely edit the social graphs based on the cosine similarity between users' collaborative representations.
\item EIISRS ~\cite{chen2024social}: EIISRS refines user social representations by reconstructing social relations using a layerwise graph-enhanced variational autoencoder.
\item RecDiff ~\cite{li2024recdiff}: RecDiff introduces a social denoising framework that directly applies diffusion probabilistic models in the hidden space to refine encoded user representations.
\item GDSSL ~\cite{li2024graph}: GDSSL applies the discrete-time diffusion algorithm directly to the raw social adjacency matrix to mitigate social noise, subsequently employing SSL to align cross-domain information.
\end{itemize}

For all baselines, we adhere to their official implementations and recommended settings to ensure fair comparisons. Specifically, the SocialMF, NGCF, LightGCN, DiffNet, SEPT, and MHCN were implemented using the publicly available QRec~\footnote{https://github.com/Coder-Yu/QRec} library which is built on TensorFlow. The implementations of DiffRec, DreamRec, BSPM, GiffCF, Design, DSL, GBSR, and RecDiff were derived from their official code releases. EIISRS and GDSSL were replicated following the original papers. The DreamRec+ model integrates social information obtained through the GCN encoder $\text{\textbf{SoEnc}}(\cdot)$ as the guidance embedding for DDPM, utilizing a multilayer perceptron for this integration. 
Moreover, we implement social denoising baselines requiring a recommendation backbone model, as well as our proposed SGSR with the most widely used LightGCN backbone. For a fair comparison, we employ a grid search strategy to tune the hyperparameters of the baseline models, exploring the vicinity of the optimal values reported in their respective original papers. All methods are optimized using the Adam optimizer. For the diffusion-based recommenders, we meticulously tune both the upper and lower bounds of the noise, as well as the number of diffusion steps, to maximize performance. Take 5 steps in DiffRec as an example, the upper bound and lower bound are searched in $\{$0.5, 0.1, 0.05, 0.01 $\}$ and $\{$0.01, 0.001, 0.0001, 0.0001 $\}$.

\noindent  \textit{\textbf{ (4) Parameter Settings.}}
The proposed framework is implemented with PyTorch. 
%Hyperparameters are optimized based on Recall@10. 
The embedding dimension $d$ is tuned within $\{$32, 64, 128, 256$\}$. 
% We use batch sizes of 1024 for Ciao and 2048 for Epinions and Dianping and sample one negative instance for each positive instance to compute the BPR loss. 
We use batch sizes of 1024 for Ciao and 2048 for Epinions and Dianping to improve training efficiency based on the scale of each dataset.
For each positive instance, we sample one negative instance to compute the BPR loss.
We use 3 GNN layers to encode the high-order social relations for our denoising framework.
We employ the Adam optimizer with a learning rate of 0.001 to train the SGSR model.
The diffusion step $T$ is evaluated across $\{3,5,10,30,50,100\}$ for both VE-SDE and VP-SDE formulations. 
We examine two predictors (reverse diffusion sampler, Euler-Maruyama) in combination with two correctors (Langevin MCMC, none). 
The maintain rate $\rho$ and $\hat{\rho}$ in the curriculum learning mechanism are tuned within $\{$0.7, 0.75, 0.8, 0.85, 0.9$\}$. 
Additionally, the temperature parameter for SSL, $\tau$, which is set in $\{0.05, 0.1, 0.5\}$. The loss weights $\lambda_1$ and $\lambda_2$ are tuned in $\{0.001, 0.01, 0.1, 0.5, 1, 5\}$. 
% The noise scale for the 5-step diffusion process is $\{0.05, 0.5\}$ after the experiment.

\begin{figure}[tp]
\centering
\subfloat[]{\includegraphics[width=1.72in]{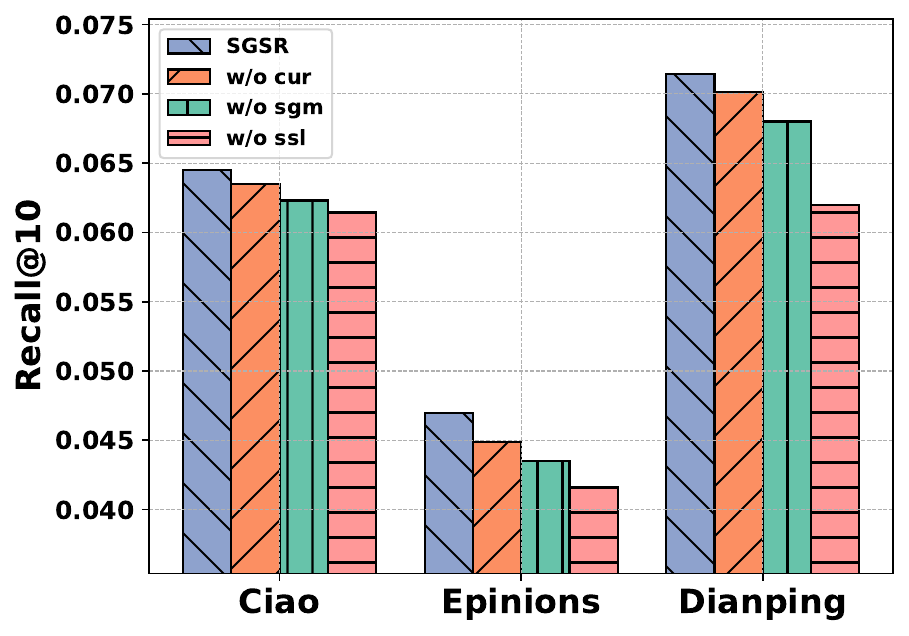}%
\label{fig:ablation_recall}}
\hfil
\subfloat[]{\includegraphics[width=1.72in]{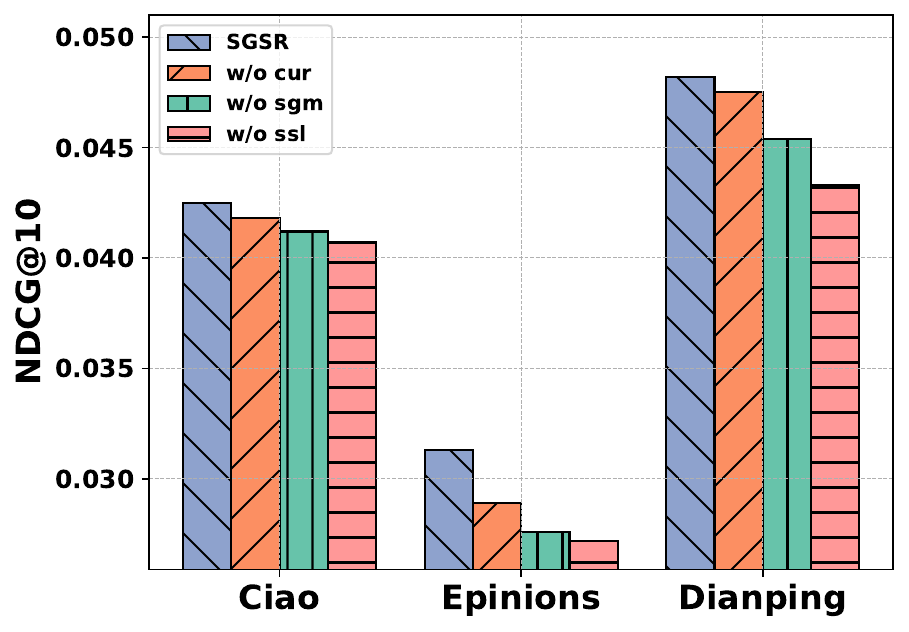}%
\label{fig:ablation_ndcg}}
\caption{Comparison of SGSR with its variants on three datasets across metric Recall@10 and NDCG@10.}
\vskip -0.2in
\label{fig:ablation}
\end{figure}

\subsection{Overall Performance}
We evaluate the recommendation performance of all baselines and our proposed approach SGSR. Table \ref{OverallPerfromance} summarizes the Recall and NDCG score across the Ciao, Epinions, and Dianping datasets, while Table \ref{Perfromance} demonstrates varying top-$K$ performance on the Epinions datasets, specifically evaluating social recommendation modes and graph denoising methods.
%, with all results averaged over five repeated experiments. 
Our main findings include:
\begin{itemize}[leftmargin=*]
\item SGSR consistently outperforms all baseline methods, demonstrating significant improvements across all datasets. Notably, on the Ciao dataset, SGSR achieves a 5.49{\%} improvement over the top-performing baseline. This substantial performance gain highlights the effectiveness of addressing low homophily challenges from a generative perspective. In particular, the classifier-free SGM used in SGSR effectively generates collaborative-consistent social representations, which significantly contributes to its superior performance.

\item In most cases, social denoising methods surpass the GNN-based social recommendation that directly utilizes the raw social graph. This observation underscores the importance of addressing the challenge of low social homophily. 
While generative denoising methods obtain similar performance to graph editing techniques, SGSR shows improved performance over both. 
This performance gap is likely rooted in: (1) reliance on simple, handcrafted similarity metrics and restricted editing scope of graph editing methods, and (2) the inherent limitations of generative paradigms and their training paradigms in the context of social recommendation.

% \jh{For ``stems from'', how about some less GPT-like words such a ``rooted in'', ``comes from'', ``due to'', ``because of'', and ``attributed to''? I also suggest using ``obtain'' to replace ``attain''.}
% \jh{For graph editing methods, they are weak because of selecting social connections based on simple and intuitive handcrafted similarity metrics. For generative methods, I think a better way is to highlight the advantage of our generative model in a more concrete way: 1) flexibility of SGM framework 2) Our guided training process}

%In most cases, social denoising methods surpass the GNN-based social recommendation that directly utilizes the raw social graph. This observation underscores the importance of addressing the challenge of low social homophily. However, the performance of these social recommendation methods is sub-optimal compared with the proposed SGSR, likely due to their reliance on simple, handcrafted similarity metrics and the restricted scope of rewired edges. 

\item Incorporating social information typically enhances the recommendation performance of both GNN-based models and diffusion-based approaches. However, the performance of DiffNet and DreamRec+ degrades when social networks are integrated. This suggests the presence of low social homophily and indicates the necessity to filter out redundant information from social relations. 

\item Diffusion-based methods exhibit performance comparable to graph-based recommendation models, highlighting the promising potential of diffusion algorithms for recommendation tasks. Nevertheless, these methods still generally trail behind most vanilla social recommendation approaches.
% Diffusion-based methods achieve comparable performance compared with the graph-based recommendation models, demonstrating the great potential of the diffusion algorithm in the task recommendation. However, such methods underperform most vanilla social recommendation models.Diffusion-based methods underperform most vanilla social recommendation models.
This observation indicates that existing diffusion models for recommendation cannot be simply adapted to the social recommendation task, necessitating the need to address the unique challenges in developing a diffusion-based recommender model for social recommendations.
\end{itemize}

\subsection{Ablation Study}
This study aims to evaluate the impact of the key components in SGSR. To comprehensively analyze the effectiveness of each module, we design three model variants:

\begin{table} [bp]
\vskip -0.1in
\centering
\caption{Ablation Study on Generative Models}
% \vskip -0.15in
\begin{tabular}{c|cc|cc}
\toprule
\toprule 
\textbf{Dataset} & \multicolumn{2}{c|}{\textbf{Ciao}} & \multicolumn{2}{c}{\textbf{Epinions}} \\
\cmidrule{1-5}
\textbf{Method} & \textbf{R@10} & \textbf{N@10} & \textbf{R@10} & \textbf{N@10} \\
\hline
SGM (VP-SDE) (Ours)        & \textbf{0.0645} & \textbf{0.0425} & \textbf{0.0470} & \textbf{0.0315} \\
SGM (VE-SDE) (Ours)        & 0.0630     & 0.0419   & 0.0453     & 0.0295\\
SGM (VP-SDE w/o predictor) & 0.0614     & 0.0416   & 0.0420     & 0.0303\\
DDPM                       & 0.0615     & 0.0415   & 0.0417     & 0.0290\\
NCSN                       & 0.0611     & 0.0412   & 0.0408     & 0.0287\\
VAE                        & 0.0607     & 0.0411   & 0.0401     & 0.0282\\

\bottomrule
\bottomrule

\end{tabular}
\label{tab:ablation_generative}
% \vskip -0.15in
\end{table}

\begin{itemize}[leftmargin=*]
\item \textit{w/o cur}: This variant removes the Curriculum Learning Mechanism, allowing the SGSR to directly learn the social representation from the raw social graph ${\mathcal{G}}^{S}$. Consequently,  $L^{Diff}$ is reduced to Eq.(\ref{eq:SDE_loss_dis}).
\item \textit{w/o sgm}: This variant eliminates the Score-based Generative Model (SGM), instead employing traditional data augmentation techniques to generate two social views for cross-domain contrastive learning. 
% (the convergence analysis is provided in Appendix ~\ref{sec:convergence})
\item \textit{w/o ssl}: This variant replaces the contrastive learning module with multi-layer perceptrons to fuse user representations from the social and collaborative domains.
\end{itemize}

Figure \ref{fig:ablation} presents the results of the ablation study. Firstly, the removal of the SGM resulted in significant performance degradation, underscoring its crucial role in the overall architecture. This finding highlights the importance of SGM in enhancing the model's predictive capabilities. Secondly, we observed a reduction in performance when the curriculum training module was eliminated. This decline suggests that the inclusion of the entire raw social graph can negatively impact the model's effectiveness, demonstrating the importance of social denoising tasks. The curriculum training approach appears to mitigate these effects by providing a more structured learning process. Lastly, the most substantial performance decrease was observed upon the removal of the Self-Supervised Learning (SSL) component. 
This outcome aligns with our previous experimental results and strongly indicates the effectiveness of the contrastive learning module. The SSL component plays an important role in capturing nuanced features and relationships within the data, thereby significantly contributing to the model's overall performance.

\begin{table} [tp]
% \vskip -0.15in
\centering
\caption{Effect of Different Encoders}
% \vskip -0.15in
\begin{tabular}{cc|cc|cc}
\toprule
\toprule 
\multicolumn{2}{c|}{\textbf{Dataset}} & \multicolumn{2}{c|}{\textbf{Ciao}} & \multicolumn{2}{c}{\textbf{Epinions}} \\
\cmidrule{1-6}
\textbf{UI-Encoder} & \textbf{Social-Encoder} & \textbf{R@10} & \textbf{N@10} & \textbf{R@10} & \textbf{N@10} \\
\hline
\textbf{LCN} & \textbf{GCN} & 0.0645 & 0.0425 & 0.0470 & 0.0315\\
LCN & LCN & 0.0647 & 0.0422 & 0.0469 & 0.0316\\
GCN & GCN & 0.0641 & 0.0419 & 0.0465 & 0.0314\\
GAT & GAT & 0.0636 & 0.0413 & 0.0459 & 0.0310\\
\bottomrule
\bottomrule

\end{tabular}
\label{tab:ablation_encoders}
\vskip -0.15 in
\end{table}

\noindent \textit{\textbf{Ablation Study on Generative Models.}} We further conduct comparison experiments with alternatives of generative models including VAE, DDPM ~\cite{ho2020denoising}, NCSN ~\cite{song2019generative}, and different variants of SGMs. For VAE, DDPM, and NCSN, we used their official code to ensure a fair comparison. For the SGM variants, the implementation details are provided in Section ~\ref{sec:Diffusion_Steps}, achieving different instantiations of SGM with various drift and diffusion coefficients. The results are summarized in the Table ~\ref{tab:ablation_generative}. 

From the table, we observe the following: (1) Diffusion-based generative models outperform VAE, likely due to diffusion algorithms breaking down denoising tasks into multiple steps, which enhances effectiveness. (2) DDPM achieves results comparable to SGM (VP-SDE without predictor), as DDPM can be seen as a discretized version of SGM ~\cite{song2020score}. (3) Variants of SGM outperform both DDPM and NCSN, demonstrating the effectiveness of the PC sampler and justifying the advantage of SGM.

Besides, GANs are not included in this table. As mentioned earlier, generative models like GANs are not directly applicable to social denoising. Due to the challenge of dealing with unknown optimal social networks, training GANs in this context is infeasible, as it becomes difficult to generate suitable positive and negative samples for effective training.

\begin{figure}[bp]
\vskip -0.2in
\centering
\subfloat[]{\includegraphics[width=1.16in]{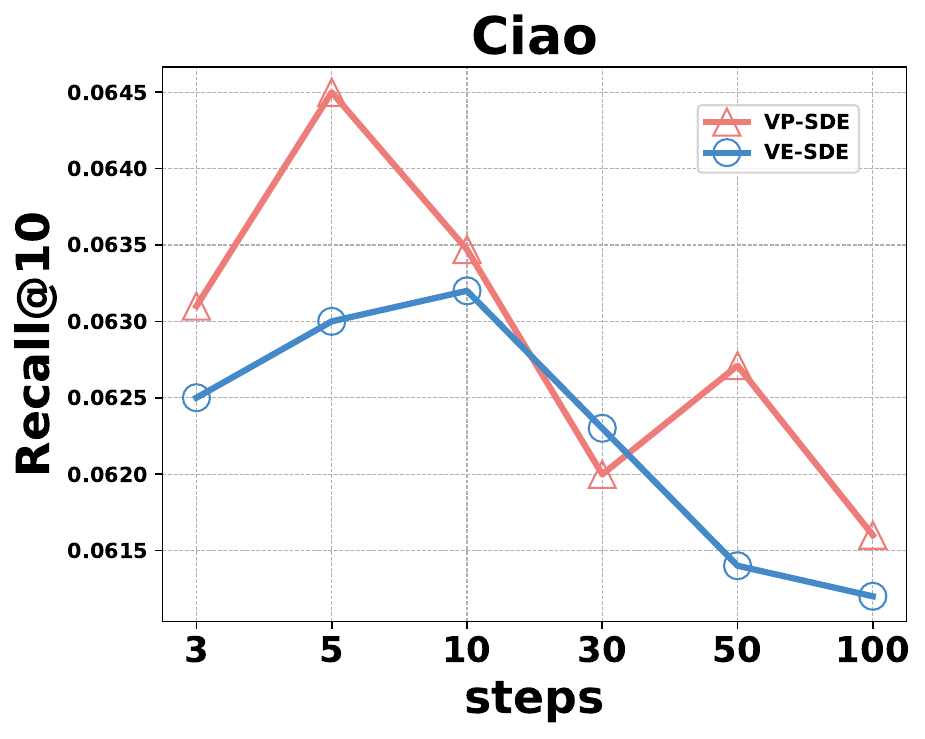}%
\label{fig:DF1}}
\hfil
\subfloat[]{\includegraphics[width=1.16in]{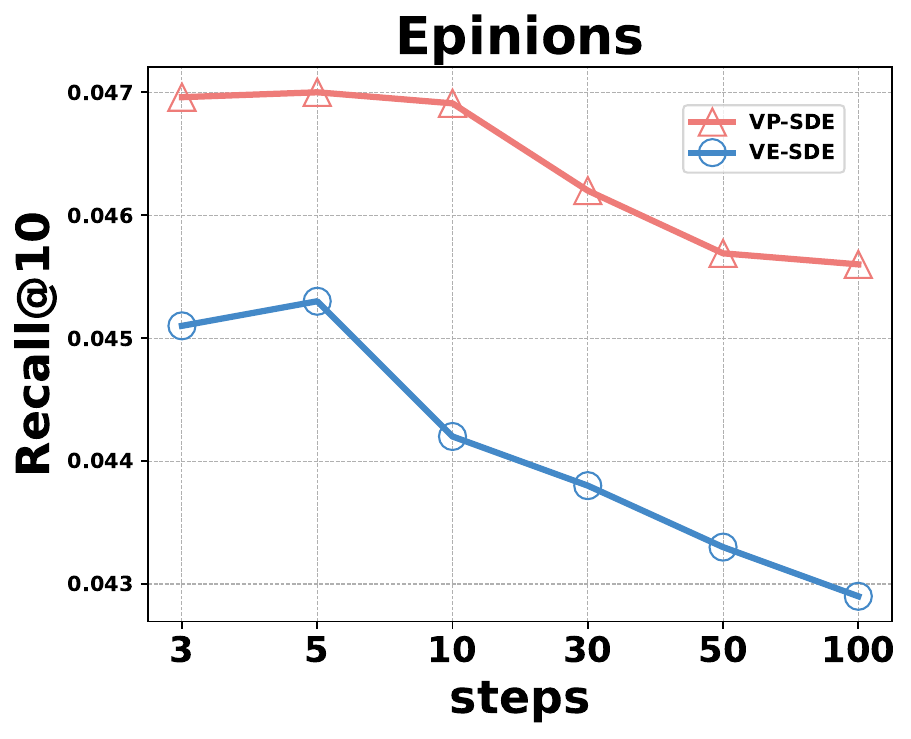}%
\label{fig:DF2}}
\hfil
\subfloat[]{\includegraphics[width=1.16in]{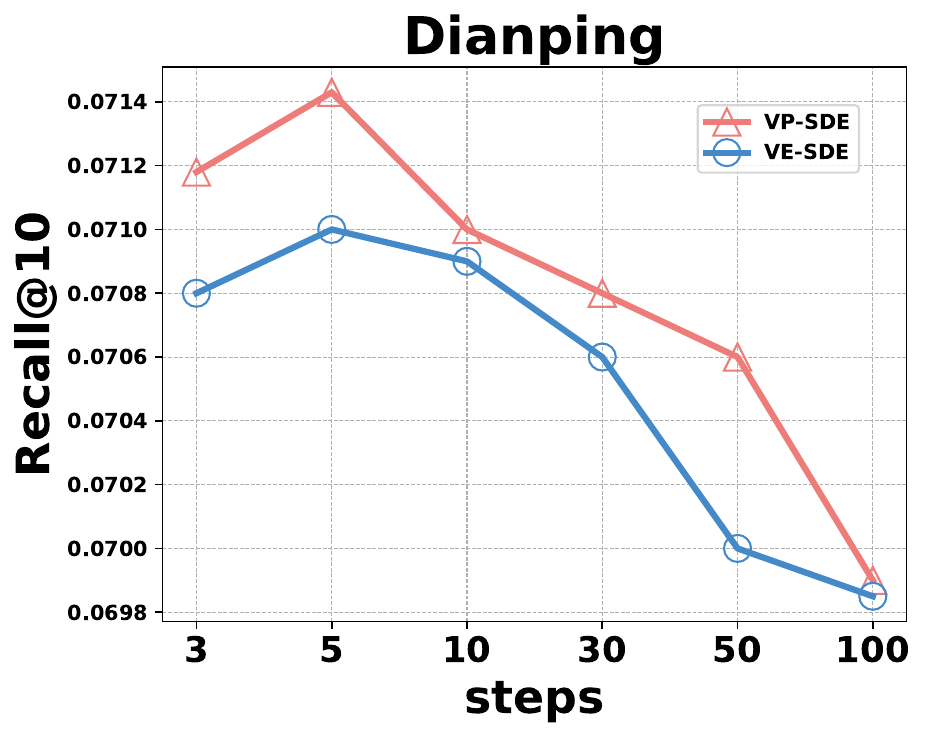}%
\label{fig:DF3}}
\hfil
\subfloat[]{\includegraphics[width=1.16in]{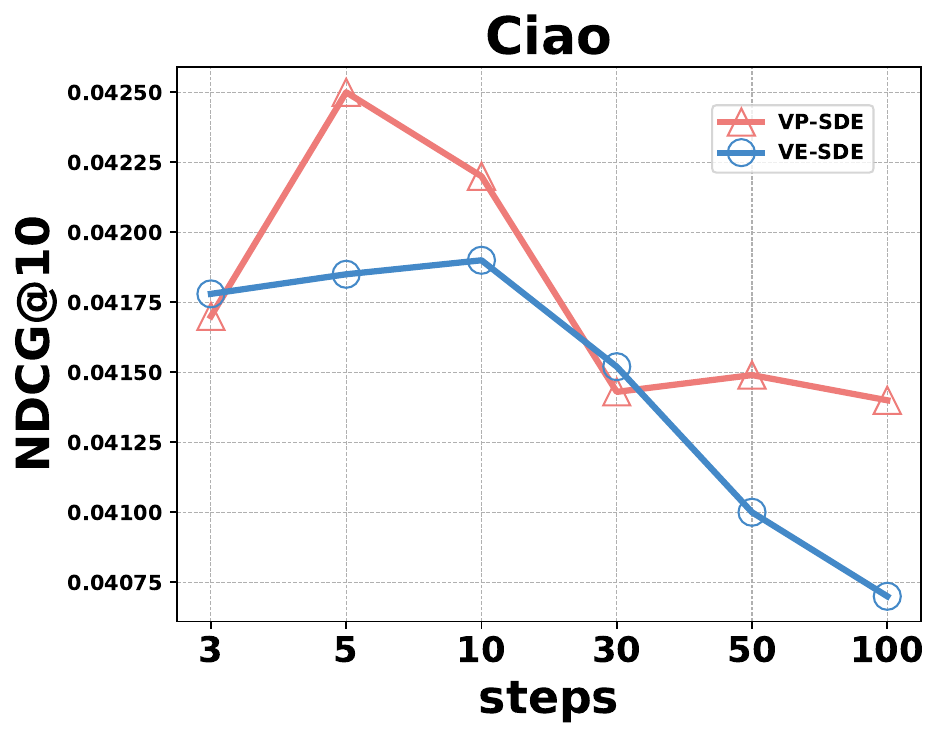}%
\label{fig:DF4}}
\hfil
\subfloat[]{\includegraphics[width=1.16in]{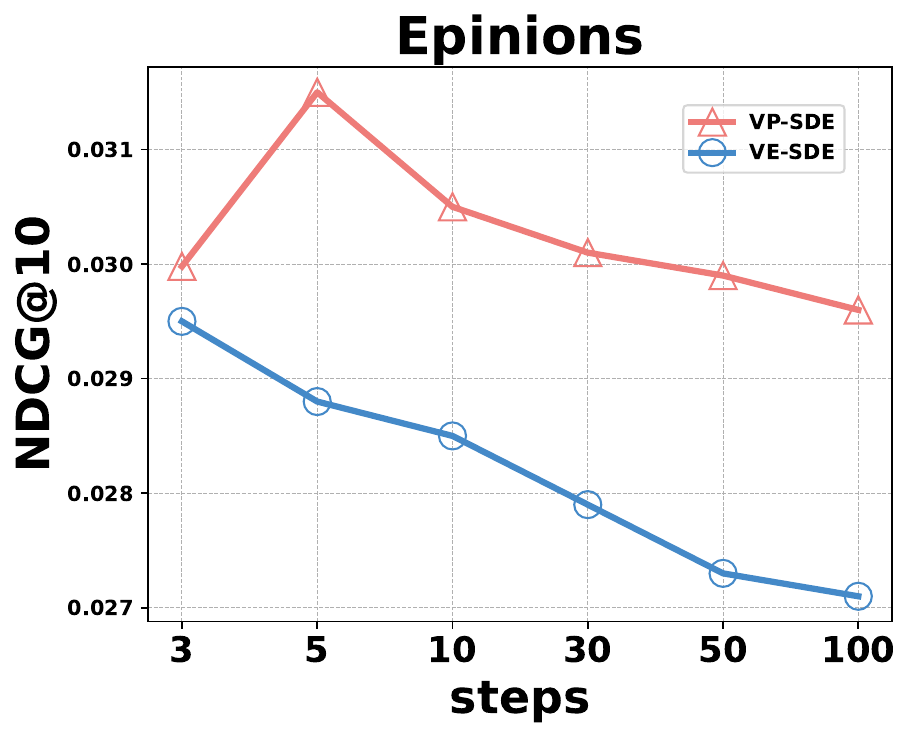}%
\label{fig:DF5}}
\hfil
\subfloat[]{\includegraphics[width=1.16in]{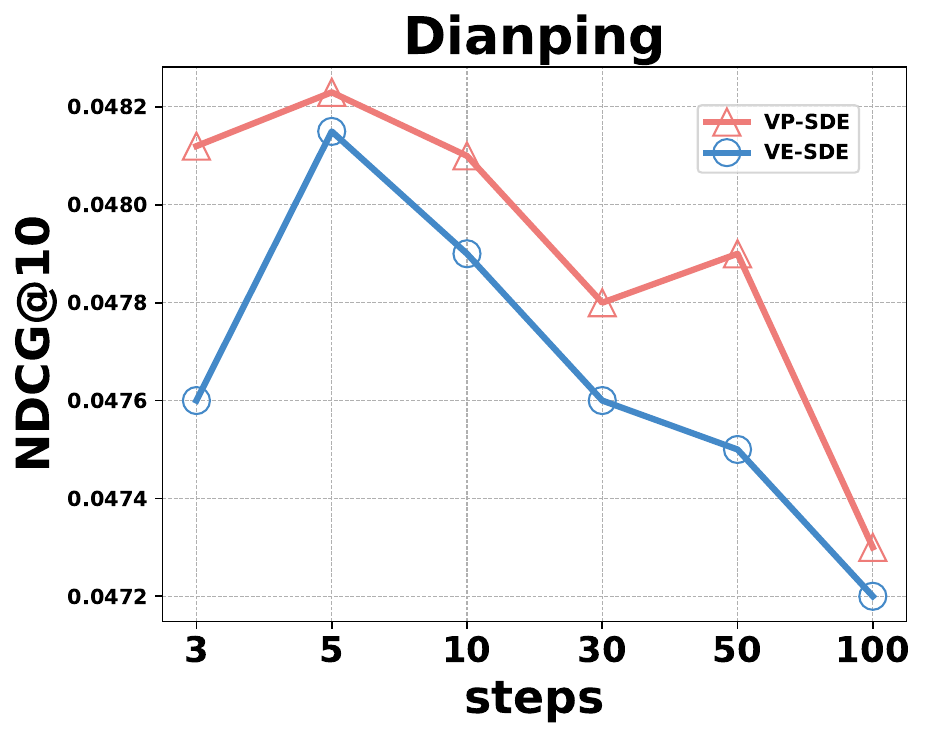}%
\label{fig:DF6}}
\caption{Effect of diffusion steps $T$ on three datasets across metrics on Recall@10 and NDCG@10.}
\label{fig:diffusion_steps} 
\vskip -0.13in
\end{figure}

\noindent \textit{\textbf{Ablation Study on Graph Encoders.}}
Graph encoders are employed to extract collaborative and social information from graph-structured data. We evaluate SGSR with various recommendation backbones to assess the impact of different encoders, following the designs of previous works. The results, presented in Table ~\ref{tab:ablation_encoders}, show that the proposed method achieves comparable performance with mainstream encoders. SGSR is adaptable to different encoders, even though each encoder produces representations of varying quality. This can be attributed to SGM, which enhances the denoised social representation with the learned data distribution.

% Our proposed SGSR framework is adaptable to various encoders. The tables ~\ref{tab:ablation_encoders} illustrate the impact of various encoders, following the designs used in previous works. Different encoders can achieve comparable results, with variations likely attributable to the specific characteristics of the graphs and encoders.

\subsection{Parameter Sensitivity Study}
In this section, we evaluate the impact of the key parameters in SGSR.

\noindent \textit{\textbf{ (1) Diffusion Steps.}}
% \subsubsection{Diffusion Steps}
\label{sec:Diffusion_Steps}
We tune the scale of the continuous time $T$, known as diffusion steps, on two well-established formulations: VP-SDE and VE-SDE. The results are presented in Fig. \ref{fig:diffusion_steps}. For all the datasets, we observe optimal performance at 5 diffusion steps. Performance degrades as the number of time steps increases, potentially due to the introduction of additional noise in longer diffusion processes. 
%\jh{\sout{The superior performance of $T=10$ compared to $T=3$ provides empirical support for the SGM.}} 
% Moreover, VP-SDE outperforms VE-SDE in most cases. 
Moreover, VP-SDE demonstrates superior performance to VE-SDE in most cases, with the performance gap widening when the diffusion step ranges between 5-50. This can be attributed to VE-SDE's tendency to significantly perturb the data distribution through high-variance noise ~\cite{song2020score}, potentially leading to personalized information loss and consequently increasing the difficulty of the reverse process.
% \jh{For this claim on VE-SDE's tendency, do we have citations?}
%\jh{\sout{The score-based diffusion algorithm enables experimentation with various SDEs under different scenarios, offering flexibility in model design and optimization.}}

\begin{figure}[t]
\centering
\subfloat[]{\includegraphics[width=1.72in]{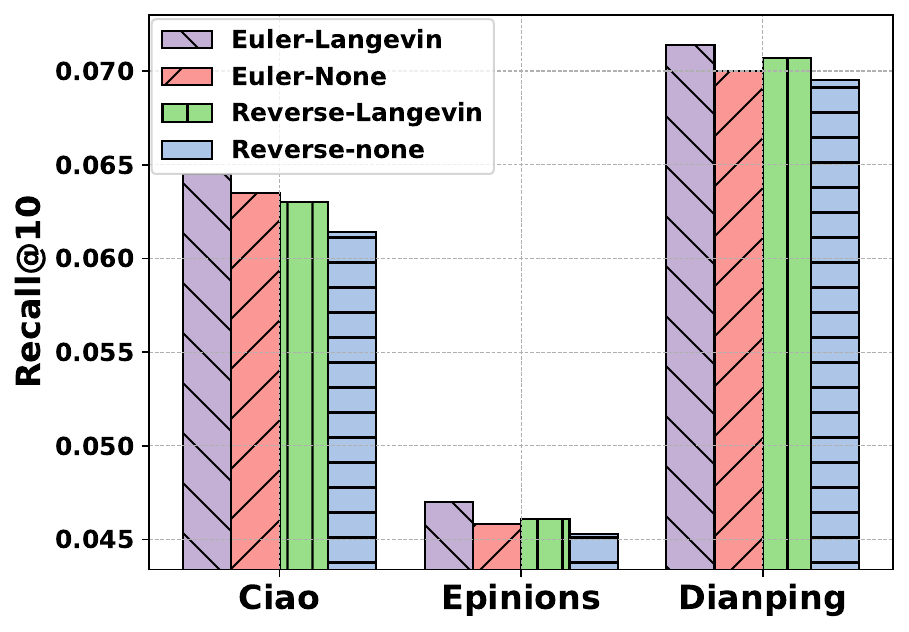}%
\label{fig:pc_recall}}
\hfil
\subfloat[]{\includegraphics[width=1.72in]{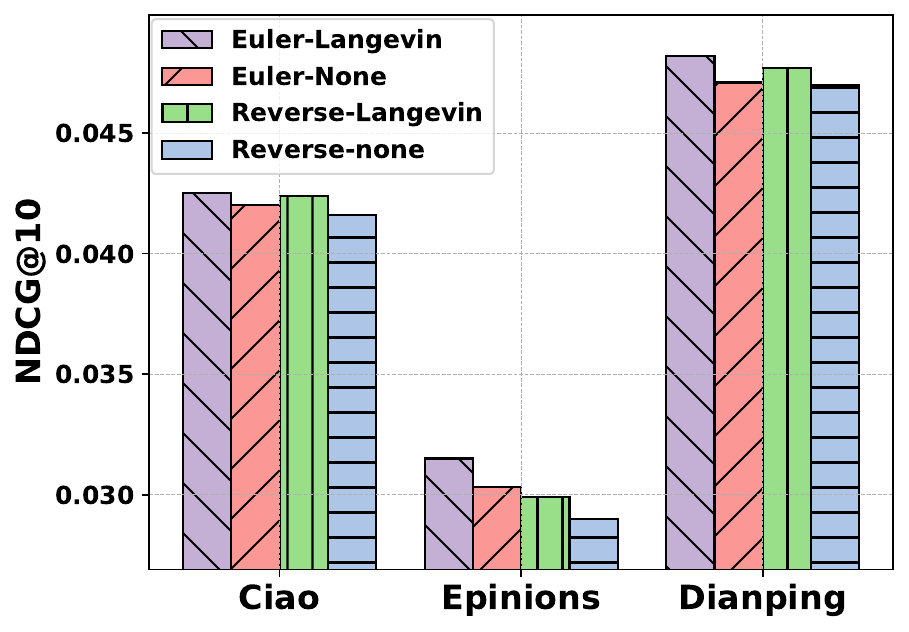}%
\label{fig:pc_ndcg}}
\caption{Effect of PC sampler on three datasets across metrics on Recall@10 and NDCG@10.}
\label{fig:pc} 
\vskip -0.25in
\end{figure}

\noindent \textit{\textbf{ (2) PC Sampler.}}
% \subsubsection{PC Sampler}
We implemented the Predictor-Corrector (PC) sampler using various combinations, including Euler-Maruyama and Langevin dynamics, Euler-Maruyama without correction, reverse diffusion sampler and Langevin dynamics, and reverse diffusion sampler without corrector. The reverse diffusion sampler is a discretization of DDPM ~\cite{song2020score}. The results, presented in Fig. \ref{fig:pc}, demonstrate that the predicting and correcting paradigm consistently outperforms the non-corrector version across all three datasets. This highlights the effectiveness of the PC sampler and the flexible sampling capabilities of SDE-based diffusion models.

\begin{figure}[thp]
\vskip -0.15in
\centering
\subfloat[]{\includegraphics[width=1.72in]{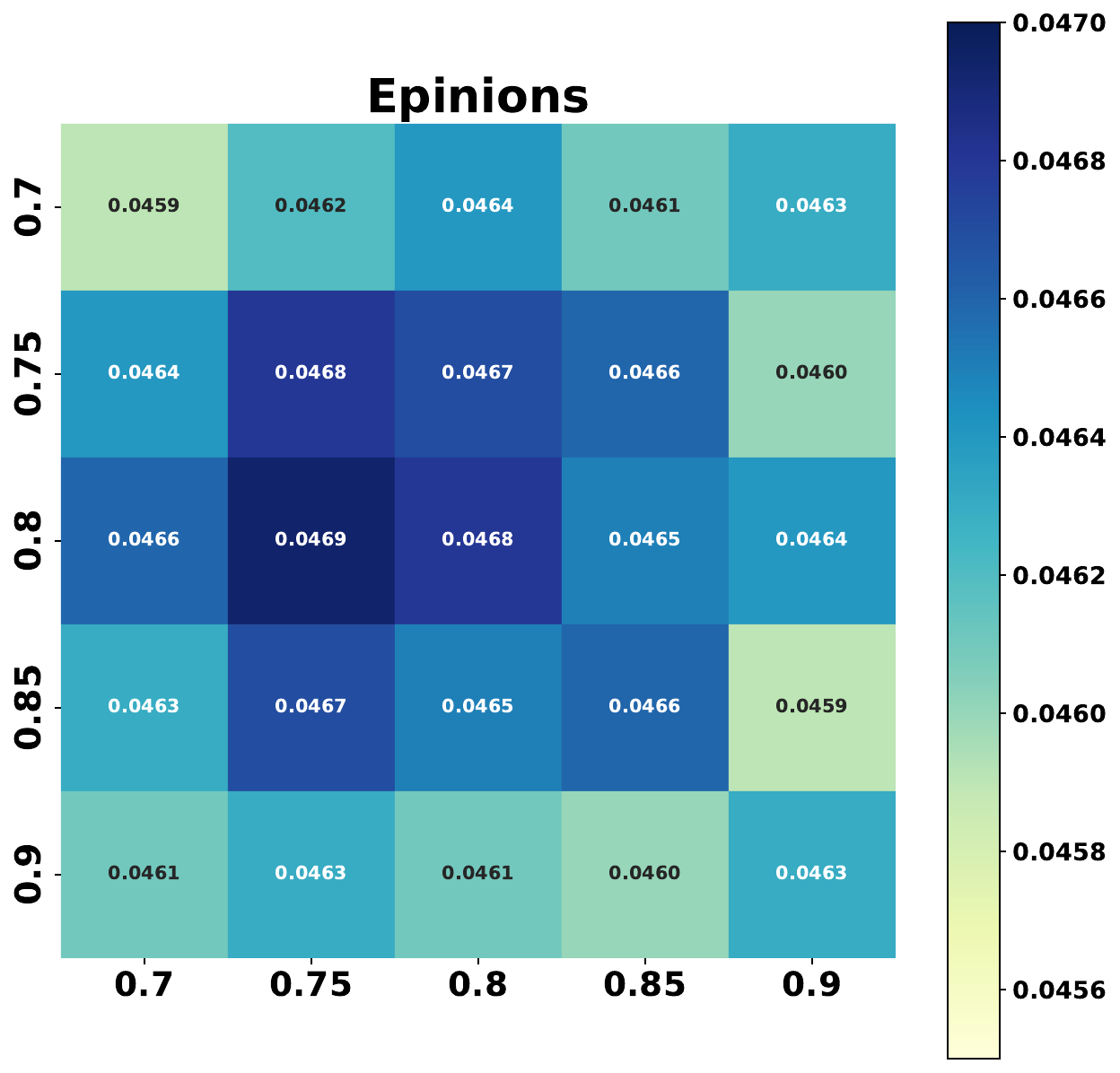}%
%\label{fig:pc_recall}
}
\hfil
\subfloat[]{\includegraphics[width=1.72in]{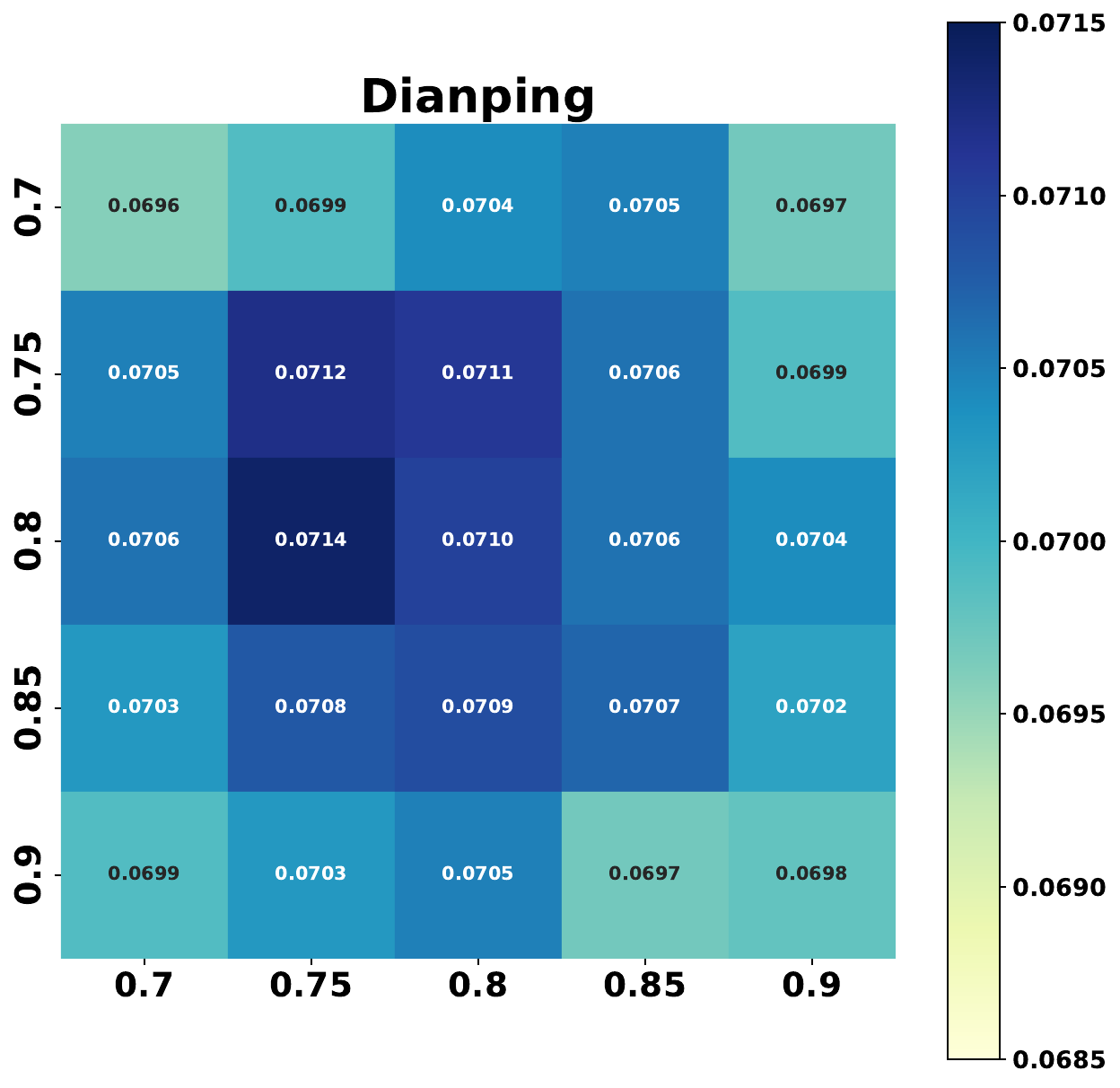}%
%\label{fig:pc_ndcg}
}
\caption{Effect of $\rho$ and $\hat{\rho}$  on metric Recall@10 for two datasets.}
\label{fig:maintain} 
\vskip -0.2in
\end{figure}

\noindent \textit{\textbf{ (3) Maintain Rate in Curriculum Learning Mechanism.}}
% \subsubsection{Maintain Rate in Curriculum Learning Mechanism}
We varied the maintain rate $\rho$ and $\hat{\rho}$ within the set $\{0.7, 0.75, 0.8, 0.85, 0.9\}$. The results are presented in Fig. ~\ref{fig:maintain}. SGSR performance peaks when we remove $20\%$ and $25\%$ of social relations for the Dianping dataset during the two-stage curriculum learning. 
This suggests that retaining approximately $80\%$ of social connections during training enables SGSR to learn optimally collaborative-consistent representations.
These results corroborate prior findings on social graph denoising ~\cite{quan2023robust, jiang2024challenging}, confirming that raw social graphs containing undesired noise connections negatively impact recommendation performance.

% This underscores the necessity of social graph denoising, as including the entire raw social graph would degrade recommendation performance. 

\subsection{Model Investigation}
In this section, we provide a deeper analysis of GBSR, focusing on memory cost, time analysis,  scalability limitation, and training efficiency.

\noindent \textit{\textbf{ (1) Memory Analysis.}}
\label{sec:Memory}
We conducted additional experiments to assess the memory consumption of the social denoising models on the Ciao and Epinions datasets, employing consistent settings (batch size: 1024, embedding size: 128, 3-layer GNN encoder for both social embeddings and collaborative embeddings). The Table ~\ref{tab:memory}, reported in MiB, demonstrates that while memory efficiency is not a particular strength of SGSR among social denoising methods, memory usage remains within acceptable bounds.
% \jh{For these complexity analysis, how about moving them to the ``Score-based Denoising Model Optimization'' Section?}

% \cy{
% Theoretically, we analyze the space complexity of SGSR, primarily including graph encoders and the reverse diffusion process. The social encoder $\text{\textbf{SoEnc}}(\cdot)$ requires $\mathcal{O}(|\mathbf{S}|+ d^2 + |U|d)$, assuming the input and output dimensions of GCN are equal, where $|\mathbf{S}|$ represents the number of social connections. Meanwhile, the collaborative encoder $\text{\textbf{RecEnc}}(\cdot)$ (e.g., LightGCN) needs $\mathcal{O}(|\mathbf{R}| + (|\mathcal{U}|+|\mathcal{V}|)d)$, with $|\mathbf{R}|$ representing the number of item interactions. 
% Aditionally, the reverse diffusion process and the scoring network $\mathbf{s}_{\boldsymbol{\theta}}$ jointly require $\mathcal{O}(dT + \sum_{\ell=1}^{\mathcal{L}}{d_{\ell-1} \cdot d_{\ell}})$ space, where $T$ is the number of diffusion step, $\mathcal{L}$ is the number of layers in the scoring network $\mathbf{s}_{\boldsymbol{\theta}}$, and $d_{\ell}$ represents the dimension of layer ${\ell}$'s output.
% }
% \jh{How about assuming that all the dimensions $d_1, \ldots,d_\mathcal{L}$ equals to $d$, or they are just at the same order as $d$? This can simplify the results and make it clearer. BTW, for the notation of number of layers, do we use $\mathcal{L}$ or $L$? We have already used $\mathcal{L}$ for loss functions.}

\begin{table}[bp]
\centering
\vskip -0.1in
\caption{Memory Analysis}
% \vskip -0.15in
\begin{tabular}{c|cc}
\toprule
\toprule 
\multirow{2}{*}{\textbf{Method}} & \multicolumn{2}{c}{\textbf{Maximum Memory Requirement (MiB)}} \\
% \cline{2-3}
 & \textbf{Ciao} & \textbf{Epinions} \\
\hline
DSL   & 2,942 & 13,949 \\
GBSR  & 3,814 & 24,330 \\
SHaRe & 3,101 & 11,993 \\
SGSR  & 3,299 & 15,643 \\
\bottomrule
\bottomrule
\end{tabular}
\label{tab:memory}
% \vskip -0.15in
\end{table}

\begin{table}[thp]
% \vskip -0.15in
\centering
\caption{Time Analysis}
% \vskip -0.15in
\begin{tabular}{c|ccc}
\toprule
\toprule 
\multirow{2}{*}{\textbf{Method}} & \multicolumn{3}{c}{\textbf{Training Time}} \\
% \cmidrule{1-2} 
 & \textbf{Ciao} & \textbf{Epinions} & \textbf{Dianping}  \\
% & \multicolumn{2}{c}{\textbf{Training Time}} \\
% \cmidrule{2-3}
\hline
DESIGN & 7.1s & 17.8s  & 57.6s\\
DSL    & 3.9s & 9.5s   & 39.1s\\
GBSR   & 6.7  & 17.2s  & 56.9s\\
SHaRe  & 5.3s & 13.7s  & 47.2s\\
% \cy{GDSSL} & 5.7s & 17.1s  & 52.9s \\
SGSR   & 7.6s & 18.3s  & 57.1s\\
\bottomrule
\bottomrule

\end{tabular}
\label{tab:time}
\vskip -0.13in
\end{table}

\noindent \textit{\textbf{ (2) Time Analysis.}}
\label{sec:Time}
We record the training time for various social denoising methods on three datasets, using consistent hyperparameter settings (batch size: 1024, embedding size: 128, learning rate: 0.0001, optimizer: Adam, 3-layer GNN encoder for both social embeddings and collaborative embeddings). The results are presented in Table~\ref {tab:time}. Despite leveraging the 5-step diffusion process, SGSR's time consumption remains high due to the stepwise denoising. Besides, the training of SGSR learns the distribution of social representations rather than merely estimating user similarity. Future work may focus on achieving one-step embedding generation with a consistent model to reduce time costs.
% The table ~\ref{tab:time} shows the training time on Ciao. Despite using 5-step diffusion process, SGSR's time consumption remains high due to the stepwise denoising. Future work may focus on achieving one-step generation with a consistent model to reduce time costs.

% \cy{
% From a theoretical perspective, the time complexity analysis of SGSR primarily consists of two components: GNN-based representation aggregation and diffusion-based denoising process. Initially, SGSR utilizes GCN to produce social embeddings. Each layer involves message passing and feature aggregation operations, yielding a time complexity of $\mathcal{O}(|\mathbf{S}|d + |\mathcal{U}|d^2)$.
% % where $|\mathbf{S}|$ represents the number of social connections. 
% The time complexity of $\text{\textbf{RecEnc}}(\cdot)$ depends on the specific GNN encoder. Taking LightGCN as an example, it requires $\mathcal{O}(|\mathbf{R}|d + (|\mathcal{U}|+|\mathcal{V}|)d)$.
% % where $|\mathbf{R}|$ represents the number of item interactions.
% Subsequently, for each batch, the forward diffusion process requires $\mathcal{O}(Bd)$ calculations to perturb the social representation, while the one-step inference costs $\mathcal{O}(B\cdot \sum_{\ell=1}^{\mathcal{L}}{d_{\ell-1} \cdot d_{\ell}})$, where $B$ denotes the batch size. 
% $\mathcal{L}$ is the number of layers in the scoring network $\mathbf{s}_{\boldsymbol{\theta}}$, and $d_{\ell}$ represents the dimension of layer ${\ell}$'s output.
% }

\noindent \textit{\textbf{ (3) Scalability Analysis.}}
\label{sec:Scalability}
In SGSR, the SGM is applied to the social representations of users, initially learned by GNN-based encoders. Unlike diffusion methods on graphs, the computational cost of the SGM does not increase with the graph size ~\cite{zhao2024denoising}. Therefore, the scalability of SGSR is not constrained by graph size, particularly due to the continuous-state diffusion model. This also motivates the idea of addressing the low-homophily challenge from a \textbf{generative} perspective rather than graph editing methods.

\begin{figure}[hbp]
\vskip -0.25in
\centering
\subfloat[]{\includegraphics[width=1.72in]{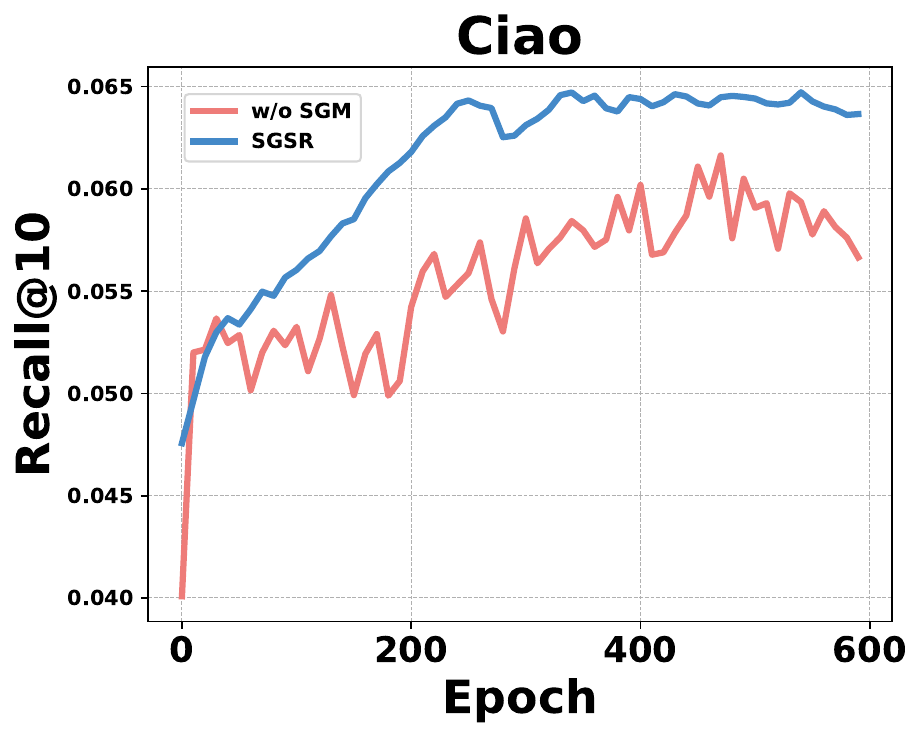}%
\label{fig:cog_recall}}
\hfil
\subfloat[]{\includegraphics[width=1.72in]{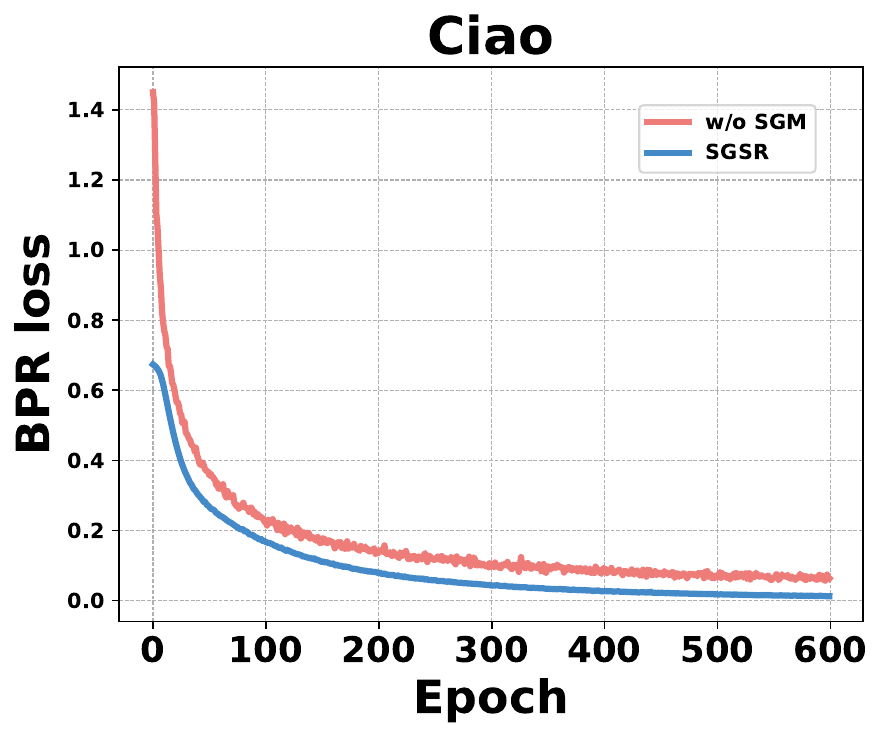}%
\label{fig:cog_bpr}}
\caption{Convergence curves of GBSR and \textit{w/o sgm} variant with respect to Recall metrics and BPR loss.}
\label{fig:converage}
\vskip -0.15in
\end{figure}

\noindent \textit{\textbf{ (4) Convergence Analysis.}}
\label{sec:convergence}
To further investigate the training efficiency of the proposed method, we compare the convergence process of SGSR and the \textit{w/o sgm} variant in terms of recall metrics and the BPR loss. 
As shown in Fig. ~\ref{fig:cog_recall}, the proposed SGSR exhibits significantly faster and more stable convergence, peaking around 350 epochs. In contrast, the \textit{w/o sgm} variant, which directly utilizes the raw social representation, converges around 500 epochs and presents considerable fluctuations.
Regarding the BPR loss trend in Fig. ~\ref{fig:cog_bpr}, the \textit{w/o sgm} variant shows notable fluctuations in loss reduction, whereas the SGSR framework demonstrates a smoother decrease.
The stability of the training process can be attributed to the effective SGM-based denoising method and the efficiency of the curriculum training algorithm.

\subsection{Statistical Analyses}

To further validate the effectiveness of SGSR, we statistically analyze the learned user representations from the Ciao dataset to evaluate their capacity for capturing real social connections.
Specifically, we compute pairwise cosine similarities between user embeddings and select the top-$K$ most similar users as predicted social connections. These predictions are then quantitatively assessed against first-order ground-truth relationships using Recall@$K$ (R@$K$) and Hit Ratio@$K$ (HR@$K$) metrics, where $K$ is in \{20, 10,  5\}.
For comparative analysis, we examine representations from GDSSL, which combines graph diffusion with self-supervised learning for social denoising, alongside LightGCN as a pure collaborative filtering baseline. 
Table~\ref{tab:Statistical_Analyses} demonstrates that both social denoising models significantly outperform LightGCN, with SGSR achieving consistent improvements over GDSSL. These results confirm that: (1) social denoising models effectively learn social connection patterns, and (2) SGSR's denoised representations better preserve social relationship structures than GDSSL's approach.

\begin{table}[hbt]

\caption{Statistical Analyses of Social Connections}
\centering
\resizebox{0.98\linewidth}{!}{
% \resizebox{0.9\linewidth}{!}{
% \setlength{\tabcolsep}{1.5mm}{
\begin{tabular}{c|ccc|ccc}
\toprule
\toprule
% \textbf{Dataset}  & \multicolumn{6}{c}{\textbf{Epinions}} \\
% \cmidrule{1-7}
\textbf{Method}   & \textbf{R@20} & \textbf{R@10} & \textbf{R@5} & \textbf{HR@20} & \textbf{HR@10} & \textbf{HR@5} \\
\hline
LightGCN & 0.0213 & 0.0133 & 0.0086 & 0.2667 & 0.1805 & 0.1996 \\
GDSSL    & 0.1866 & 0.1425 & 0.1030 & 0.7996 & 0.7056 & 0.5839 \\
SGSR     & 0.1996 & 0.1513 & 0.1122 & 0.8234 & 0.7277 & 0.6269 \\
\bottomrule
\bottomrule
\end{tabular} }
\label{tab:Statistical_Analyses}
\vskip -0.15in
\end{table}

% With the default T-SNE configuration of scikit-learn\footnote{https://scikit-learn.org/stable/modules/generated/sklearn.manifold.TSNE.html}, we visualize the learned user representations on the Ciao dataset to further validate the effectiveness of SGSR, as shown in Fig.~\ref{fig:visual}. For comparison, we include representations from LightGCN (showing pure collaborative embeddings) and GDSSL (using graph diffusion to denoise social networks with self-supervised knowledge transfer). 
% The visualization indicates that SGSR's social representations exhibit distributional patterns more closely aligned with collaborative signals than those of GDSSL, as reflected in the similarity of clustering structures. This difference can be attributed to the distinct types of information captured from the respective domains.
% These results highlight SGSR’s strong ability to directly generate collaborative-consistent social representations. In contrast, GDSSL's social embeddings exhibit greater dispersion, suggesting inherent expressive capability limitations in discrete-space optimization approaches for social connection refinement.

\section{Related Work}
\label{RelatedWork}
This section reviews related work in social recommendation and diffusion models for recommender systems.

% In this section, we summarize the related works on the social recommendation task and diffusion models for recommendations.

\subsection{Social Recommendations and Social Denoising Methods}

\noindent \textit{\textbf{ (1) General Social Recommendation Methods.}} The widespread use of online social networks has alleviated the data sparsity issue in recommender systems by incorporating social context to enhance user representation modelling. Early approaches to social recommendations adopted the social homophily assumption and directly leveraged raw social networks, evolving through matrix factorization~\cite{jamali2010matrix, ma2011recommender}, graph neural networks~\cite{wu2019neural, fan2019graph, fan2020graph}, and self-supervised learning~\cite{yu2021socially, yu2021self, xia2023disentangled}. However, these methods struggle to address the redundant or irrelevant information inherent in raw social networks, introducing noise into user representations. This problem necessitates systematic denoising methods to improve social recommendation.

\noindent \textit{\textbf{ (2) Data-centric Social Denoising Methods.}} Current social denoising methods primarily focus on modifying users’ social relations by assessing the confidence of social connections using handcrafted metrics to filter out irrelevant relations. For example, GDMSR~\cite{quan2023robust} employs a transformer layer to evaluate the confidence of social connections based on historical interactions, removing a fixed percentage of edges with the lowest similarity scores. SHaRe~\cite{jiang2024challenging} and MADM~\cite{ma2024madm} use cosine similarity to assess the reliability of social relations, discarding or rewiring connections with low similarity scores and keeping reliable connections. GBSR~\cite{yang2024graph} introduces a theoretically-inspired framework to maximize the mutual information between the denoised social graph and the interaction matrix, but it only drops the links in the social network while ignoring possibly beneficial edge rewiring or injecting operations. Generally speaking, existing data-centric social denoising methods can hardly reach the optimal social representation, since the best-fitted measure of social relations remains unknown and cannot be simply designed by human heuristics. Besides, graph editing methods only improve users’ social representations indirectly, and the editing path to the optimal graph structure corresponding to the optimal social representation may be computationally infeasible. 
This highlights the necessity for innovative generative solutions to low social homophily, which is the central focus of our study.
% This shows that it is necessary to address the low social homophily issue from an innovative generative perspective, which is the key research aim of this paper.

\noindent \textit{\textbf{ (3) Contrastive Social Denoising Methods.}} Unlike data-centric methods that modify the structure of social networks to enhance user representations, another line of research focuses on directly improving users' representations through contrastive learning. 
These methods~\cite{wang2023denoised, ma2024madm} typically align user representations from different views (e.g., social or collaborative domains) to achieve a position more consistent with collaborative signals in the latent space. 
However, contrastive denoising approaches also rely on pre-defined user similarity measurements, which may not fully leverage the supervision signal from collaborative filtering, potentially falling short of reaching the optimal user representations. 
Additionally, since contrastive denoising methods adjust the training loss function to reduce noise in social networks, they are orthogonal to our proposed framework and can be adaptively integrated. 
% We plan to explore the integration of the proposed SGSR framework with various auxiliary loss functions in future work.

\subsection{Diffusion Models for Recommendations}
Diffusion models, a powerful class of generative models, have significantly reshaped various recommendation tasks with their effective generation capabilities. In collaborative filtering (CF), DiffRec~\cite{wang2023diffusion} serves as a fundamental contribution, providing accurate recommendations from a diffusion perspective. Specifically, DiffRec encodes a user’s list of interacted items into a latent vector and employs a forward-reverse process to generate missing user-item interactions in the latent space. The diffused latent vector is then decoded into predicted user-item interactions. Following the proposal of the DiffRec framework, CF-Diff~\cite{hou2024collaborative} and GiffCF~\cite{zhu2024graph} enhanced this approach by incorporating high-order relational capabilities through graph-level diffusion models. Another notable diffusion model, DreamRec~\cite{yang2024generate}, targets sequential recommendation tasks by using a conditional generation paradigm to generate target-item embeddings based on historical interaction sequences. 
Additionally, leveraging the strong generative ability of diffusion models to produce high-quality samples or representations for recommendation has shown promise in recommendations~\cite{qu2025generative, wu2023diff4rec, zhao2024denoising}. For instance, DDRM~\cite{zhao2024denoising} refines noisy implicit user-item interaction data by applying a diffusion process to user and item representations, conditioned on their initial embeddings.
% DDRM, for example, denoises noisy implicit feedback through a diffusion process on user and item representations, conditioned on their original embeddings.
% for data augmentation has shown promise in recommendations~\cite{liu2023diffusion, wu2023diff4rec, zhao2024denoising}. 
% Moreover, other tasks, such as point-of-interest (POI) recommendation~\cite{qin2023diffusion} and knowledge-aware recommendation~\cite{jiang2024diffkg}, have also benefited from the success of diffusion models. 
However, existing diffusion recommender models are not designed to integrate denoised social information into recommendations, which differs from the primary research objective of this paper. Additionally, most existing works utilize the Denoising Diffusion Probabilistic Models (DDPM)~\cite{ho2020denoising} framework, which can be seen as a specific discretized form of Score-based Generative Models (SGM)~\cite{jo2022score}, highlighting the novelty of the proposed SGSR framework.

\section{Conclusion}

In this paper, we address the low homophily challenge in social recommendation from a generative perspective using score-based diffusion models. Specifically, we implement an SDE-based diffusion process to produce collaboratively consistent social representations conditioned on user preference patterns. Recognizing that classifier guidance is unsuitable for recommendation contexts, we derive a novel classifier-free optimization objective to facilitate the training of SGMs. To address the lack of supervision signals in recommendation scenarios, we introduce a curriculum learning mechanism and a joint training strategy. Our evaluation of SGSR on three real-world datasets demonstrates significant improvements in recommendation performance compared to existing methods, validating the effectiveness of our social denoising approach.

%  \jh{I am not sure whether I am wrong. Is it common to cite some papers in the conclusion? How about putting our discussions in rebuttal to the proposed method section? We can add a subsection called discussion.}
For future work, a promising direction is to accelerate sampling by designing consistent functions~\cite{song2023consistency} tailored to the recommendation context to enable one-step inference. 
% (training time analysis in Section~\ref {sec:Time})
Additionally, exploring various SDE solvers, such as Runge-Kutta methods, could further enhance the inference efficiency by leveraging the computational flexibility of score-based diffusion models. 
From a training perspective, investigating model distillation techniques to efficiently train a student model guided by SGSR along the distribution trajectory also presents a valuable direction.
% \jh{I suggest changing ``valuable avenue'' to ``valuable direction''.}
% Furthermore, while this paper primarily focuses on existing VP-SDEs and VE-SDEs, it is worthwhile to investigate more expressive and suitable SDEs tailored to the recommendation context. 

%investigating the potential of expressive SDEs is worthwhile. 
%This paper primarily focuses on the discretization of the SGM using DDPM and SMLD. 
%Exploring diverse SDE designs tailored to the recommendation context remains a valuable endeavor.

% For future work, a promising direction is to explore multiple SDE solvers, such as Runge-Kutta methods, to enhance the \jh{inference\sout{sampling}} process by leveraging the \jh{computational} flexibility of score-based diffusion models. \jh{Besides,} this paper primarily experiments with the discretization of the Score-based Generative Model (SGM) using DDPM and SMLD. Investigating diverse SDE designs that best fit the recommendation context is a worthwhile endeavour. Additionally, the consistency models~\cite{song2023consistency} offer a potential approach for achieving fast generation, which merits further exploration. 
% \jh{[Two directions: accelerated inference (better SDE solver + consistency models) and more expressive SDEs]}

\section*{Acknowledgments}
The research described in this paper has been partially supported by the National Natural Science Foundation of China (project no. 62102335), General Research Funds from the Hong Kong Research Grants Council (project no. PolyU 15207322, 15200023, 15206024, and 15224524), internal research funds from Hong Kong Polytechnic University (project no. P0042693, P0048625, and P0051361). This work was supported by computational resources provided by The Centre for Large AI Models (CLAIM) of The Hong Kong Polytechnic University.

\bibliographystyle{IEEEtran}
\bibliography{bare_jrnl_new_sample4}

% \input{sections/appendix_P1}
% \input{sections/appendix_P}
% \input{sections/appendix_P2}
% \input{sections/appendix_P3}

%\newpage

\section{Biography Section}

\begin{IEEEbiography}[{\includegraphics[width=1in,height=1.25in,clip,keepaspectratio]{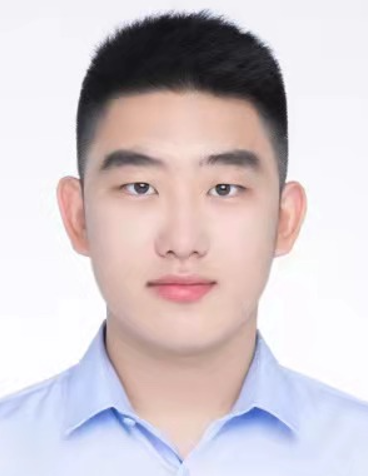}}]{Chengyi Liu} is currently a second-year PhD student at the Department of Computing (COMP), Hong Kong Polytechnic University (PolyU), under the supervision of Prof. Qing Li and Dr. Wenqi Fan. Before joining the PolyU, he received his Master’s degree in Health Data Science from the University College London (UCL) (M.Sc. in Health Data Science), under the supervision of Dr. Ken Li. In 2021, he got his bachelor’s degree in Computer Science from the University of Nottingham (B.Sc. in Computer Science). His research interests include Recommender Systems, Diffusion Models,  Graph Neural Networks, and Natural Language Processing. He has published innovative works in top-tier conferences such as IJCAI. For more information, please visit https://chengyiliu-cs.github.io/.
\end{IEEEbiography}

\begin{IEEEbiography}[{\includegraphics[width=1in,height=1.25in,clip,keepaspectratio]{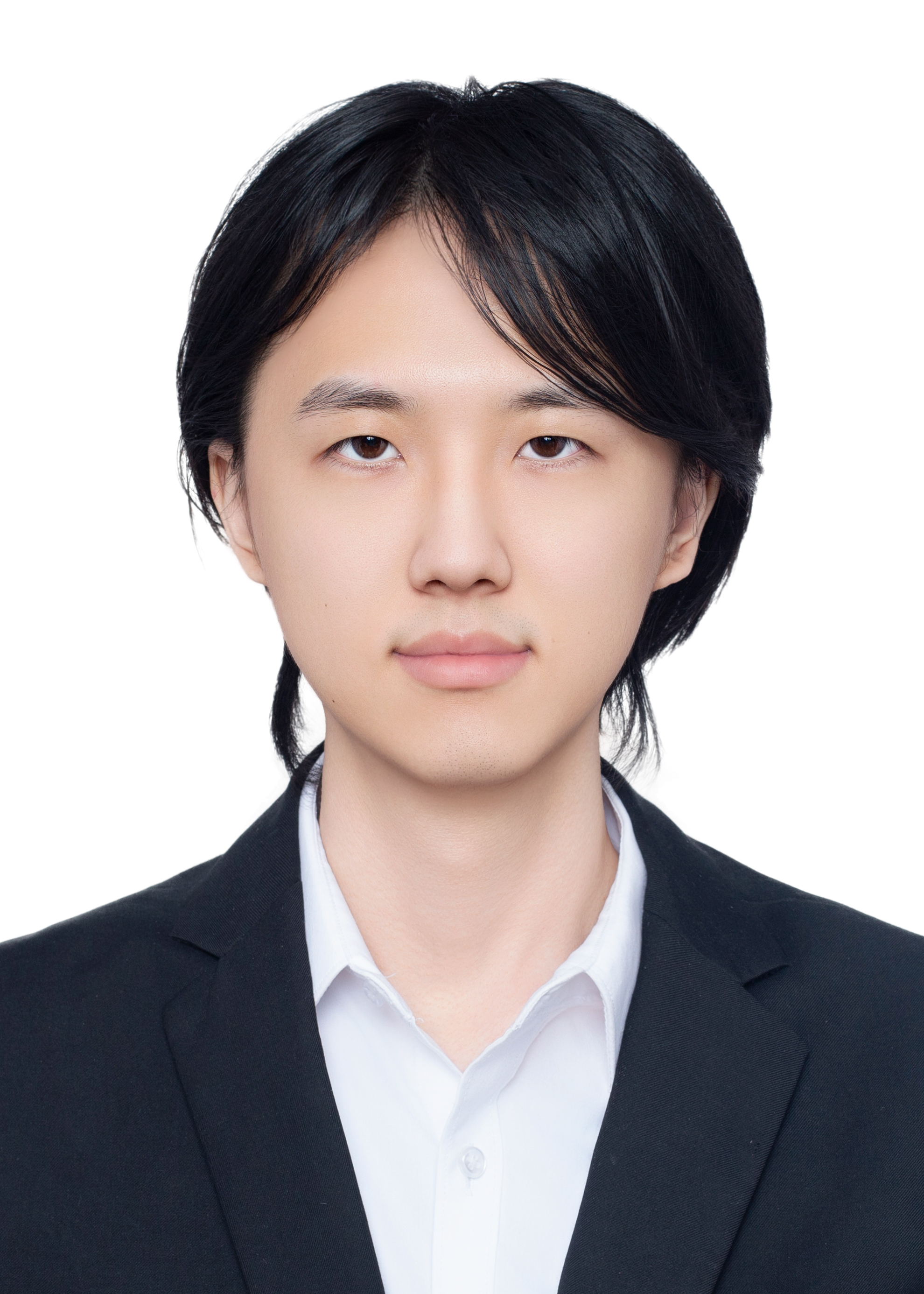}}]{Jiahao Zhang} is currently a first-year PhD student at the College of Information Science and Technology (IST), The Pennsylvania State University, under the supervision of Prof. Suhang Wang. From 2022 to 2024, he pursued an M.Phil. degree at the Department of Computing (COMP), The Hong Kong Polytechnic University, where he was jointly supervised by Prof. Wenqi Fan and Prof. Qing Li. He graduated from East China Normal University (ECNU) with a B.Eng. degree in Computer Science and received the Outstanding Graduates Award in 2022. He has published innovative works in top-tier venues such as WWW. He serves as a reviewer for top-tier conferences (e.g., KDD, WWW, AAAI, etc.) and journals (e.g., ACM TKDD, ACM CSUR, etc.). His research interest includes Graph Neural Networks, Recommender Systems, and Trustworthy Machine Learning. 
\end{IEEEbiography}

\begin{IEEEbiography}[{\includegraphics[width=1in,height=1.25in,clip,keepaspectratio]{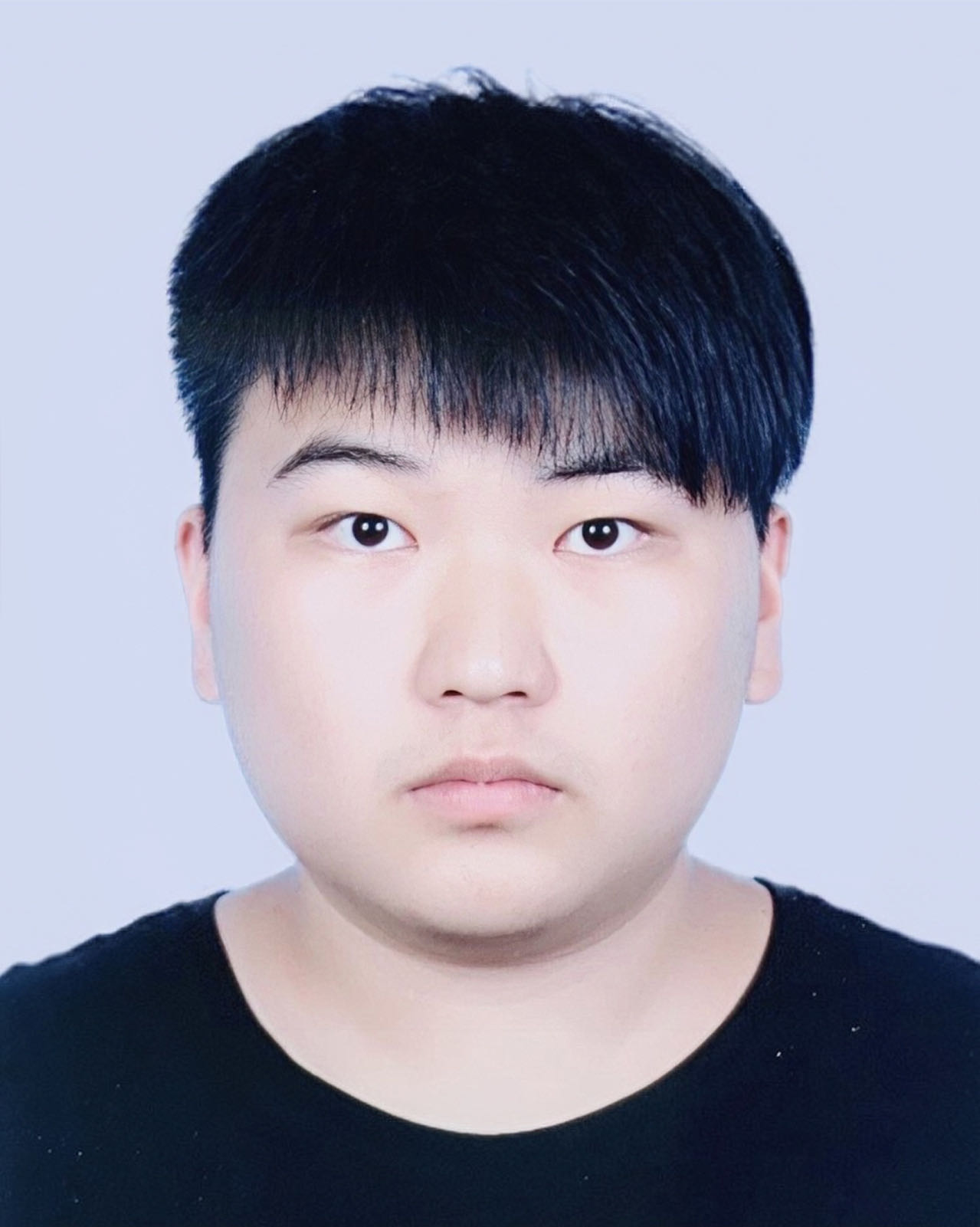}}]{Shijie Wang} is currently a third-year PhD student at the Department of Computing (COMP), Hong Kong Polytechnic University (PolyU), under the supervision of Dr. Wenqi Fan and Prof. Qing Li. Before joining the PolyU, he received his Bachelor’s degree (Hons) in Information and Computing Science from Xi’an Jiaotong-Liverpool University in China and the University of Liverpool in the UK in 2022. He serves as a top-tier conference program committee member (e.g., AAAI, ICDM, etc.), and journal reviewer (e.g., TKDD, etc.). His research interest covers Recommender Systems, Trustworthy Recommendations, and Large Language Models. He has published innovative works in top-tier journals such as TOIS. For more information, please visit https://sjay-wang.github.io/.
\end{IEEEbiography}

\begin{IEEEbiography}[{\includegraphics[width=1in,height=1.25in,clip,keepaspectratio]{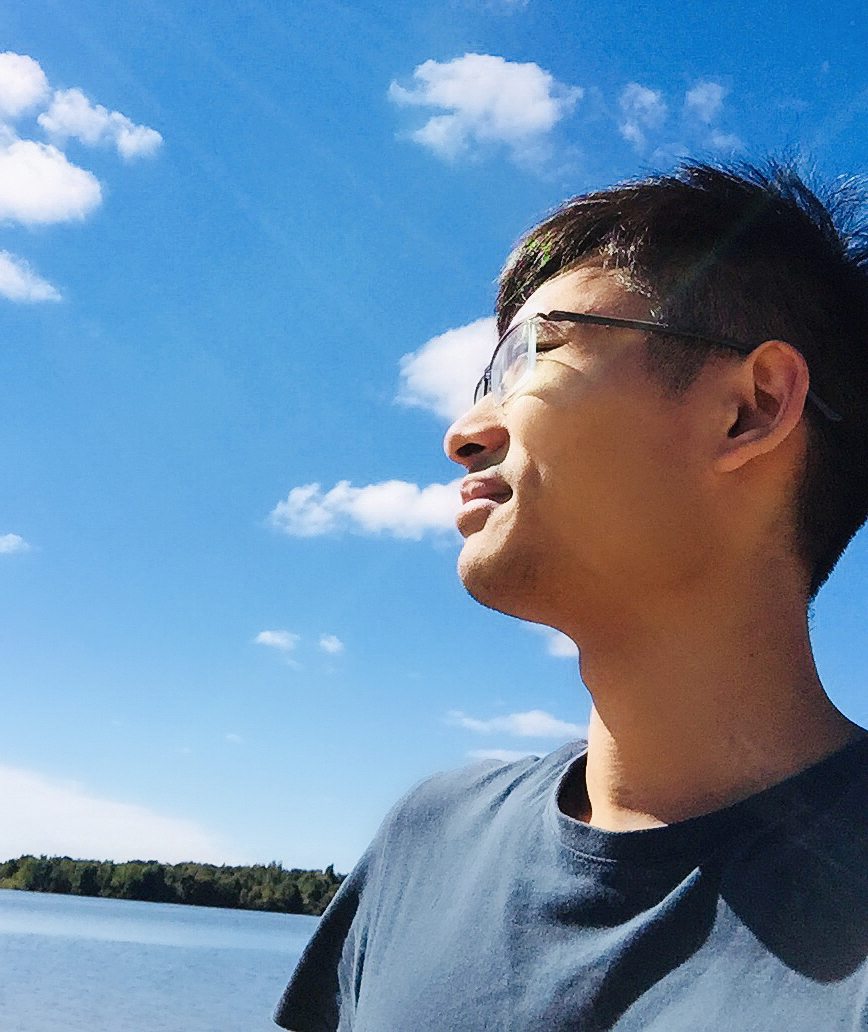}}]{Wenqi Fan} is an assistant professor of the Department of Computing (COMP) and the Department of Management and Marketing (MM) at The Hong Kong Polytechnic University (PolyU). He received his Ph.D. degree from the City University of Hong Kong (CityU) in 2020.
From 2018 to 2020, he was a visiting research scholar at Michigan State University (MSU). 
His research interests are in the broad areas of machine learning and data mining, with a particular focus on Recommender Systems, Graph Neural Networks, and Trustworthy Recommendations. He has published innovative papers in top-tier journals and conferences such as  TKDE, TIST, KDD, WWW, ICDE, NeurIPS, ICLR, SIGIR, IJCAI, AAAI, RecSys, WSDM, etc. 
He serves as a top-tier conference (Area/Senior) Program Committee member and session chairs (e.g., ICML, ICLR, NeurIPS, KDD, WWW, AAAI, IJCAI, WSDM, EMNLP, ACL,  etc.), and journal reviewers (e.g., TKDE, TIST, TKDD, TOIS, TAI, etc.). 
More information about him can be found at https://wenqifan03.github.io.
\end{IEEEbiography}

\begin{IEEEbiography}[{\includegraphics[width=1in,height=1.25in,clip,keepaspectratio]{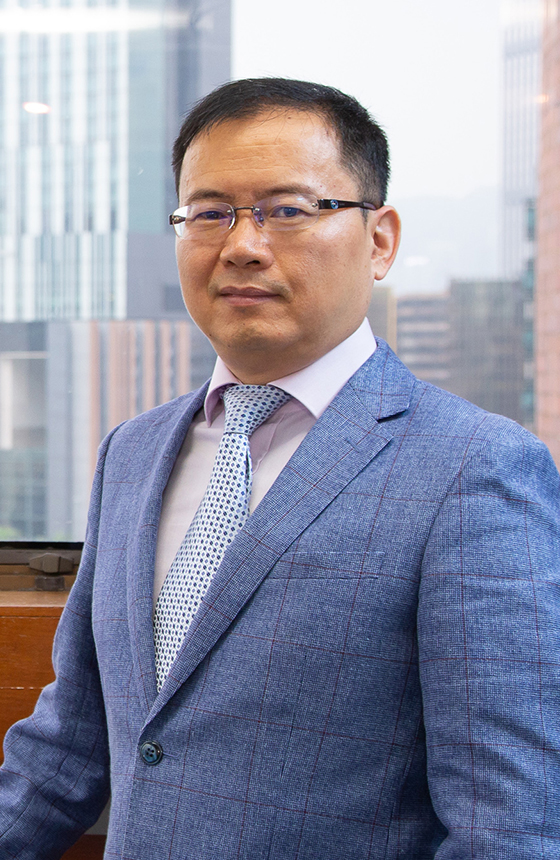}}]{Qing Li} received the B.Eng. degree from Hunan University, Changsha, China, and the M.Sc. and Ph.D. degrees from the University of Southern California, Los Angeles, all in computer science.
He is currently a Chair Professor (Data Science) and the Head of the Department of Computing, the Hong Kong Polytechnic University. He is a Fellow of IEEE and IET, a member of ACM SIGMOD and IEEE Technical Committee on Data Engineering. 
His research interests include object modelling, multimedia databases, social media, and recommender systems. 
He has been actively involved in the research community by serving as an associate editor and reviewer for technical journals and as an organizer/co-organizer of numerous international conferences. 
He is the chairperson of the Hong Kong Web Society, and also served/is serving as an executive committee (EXCO) member of the IEEE-Hong Kong Computer Chapter and ACM Hong Kong Chapter. In addition, he serves as a councillor of the Database Society of Chinese Computer Federation (CCF), a member of the Big Data Expert Committee of CCF, and a Steering Committee member of DASFAA, ER, ICWL, UMEDIA, and WISE Society.
\end{IEEEbiography}

 

% \begin{IEEEbiography}[{\includegraphics[width=1in,height=1.25in,clip,keepaspectratio]{sections/Fig/(a)DiffusionStepsCiao.pdf}}]{Michael Shell}
% \end{IEEEbiography}



\vfill

\end{document}